\documentclass[apj]{emulateapj}
\slugcomment{{\sc Accepted to ApJ:} July 10, 2006}

\def\pomega{\varpi} 
 
\def\cm{{\rm\,cm}} 
\def\m{{\rm\,m}} 
\def\km{{\rm\,km}} 
 
\def\gm{{\rm\,g}}

\def\AU{{\rm\, AU}}  
  
\def\yr{{\rm\,yr}}

\begin{document}  
  
\shortauthors{Murray-Clay \& Chiang}  
\shorttitle{Stochastic Migration}  
  
\title{Brownian Motion in Planetary Migration}  
\author{Ruth A.~Murray-Clay\altaffilmark{1} \& Eugene I.~Chiang\altaffilmark{1,2}}  
\altaffiltext{1}{Center for Integrative Planetary Sciences,  
Astronomy Department,  
University of California at Berkeley,  
Berkeley, CA~94720, USA}  
\altaffiltext{2}{Alfred P.~Sloan Research Fellow}  
  
\email{rmurray@astron.berkeley.edu, echiang@astron}  
  
\begin{abstract}  
A residual planetesimal disk of mass 10--$100M_\earth$ remained in the outer solar system following the birth of the giant planets, as implied by the existence of the Oort cloud, coagulation requirements for Pluto, and inefficiencies in planet formation. Upon gravitationally scattering planetesimal debris, planets migrate. Orbital migration can lead to resonance capture, as evidenced here in the Kuiper and asteroid belts, and abroad in extra-solar systems. Finite sizes of planetesimals render migration stochastic (``noisy'').  At fixed disk mass, larger (fewer) planetesimals generate more noise. Extreme noise defeats resonance capture. We employ order-of-magnitude physics to construct an analytic theory for how a planet's orbital semi-major axis fluctuates in response to  random planetesimal scatterings.  The degree of stochasticity depends not only on the sizes of planetesimals, but also on their orbital elements. We identify the conditions under which the planet's migration is maximally noisy. To retain a body in resonance, the planet's semi-major axis must not random walk a distance greater than the resonant libration width. We translate this criterion into an analytic formula for the retention efficiency of the resonance as a function of system parameters, including planetesimal size. We verify our results with tailored numerical simulations. Application of our theory reveals that capture of Resonant Kuiper belt objects by a migrating Neptune remains effective if the bulk of the primordial disk was locked in bodies having sizes $< \mathcal{O}(100)$ km and if the fraction of disk mass in objects with sizes $\gtrsim 1000$ km was less than a few percent. Coagulation simulations produce a size distribution of primordial planetesimals that easily satisfies these constraints. We conclude that stochasticity did not interfere with nor modify in any substantive way Neptune's ability to capture and retain Resonant Kuiper belt objects during its migration.  
\end{abstract}  
  
  
\section{INTRODUCTION}

Planet formation by coagulation of planetesimals is not
perfectly efficient---it leaves behind a residual disk of solids.
Upon their coalescence, the outer planets of our solar system were likely
embedded in a 10--$100 M_\earth$ disk of rock and ice containing
the precursors of the Oort cloud (Dones et al.~2004) and the Kuiper
belt (see the reviews by Chiang et al.~2006; Cruikshank et al.~2006; Levison et al.~2006).
The gravitational back-reaction felt by planets as they scatter and scour
planetesimals causes the planets to migrate (Fern\'andez \& Ip 1984;
Murray et al.~1998; Hahn \& Malhotra 1999; Gomes, Morbidelli, \& Levison 2004).
Neptune is thought to have migrated outward and thereby
trapped Kuiper belt objects (KBOs) into its exterior mean-motion resonances,
both of low-order such as the 3:2 (Malhotra 1995) and of high-order
such as the 5:2 (Chiang et al.~2003; Hahn \& Malhotra 2005). Likewise, Jupiter's inward migration
may explain the existence of Hilda asteroids in 2:3 resonance with the
gas giant (Franklin et al.~2004). A few pairs of extra-solar planets, locked
today in 2:1 resonance (Vogt et al.~2005; Lee et al.~2006),
may have migrated to their current locations
within parent disks composed of gas and/or planetesimals.
Orbital migration and resonant trapping of dust grains may also be required
to explain non-axisymmetric structures observed in debris disks
surrounding stars 10--100 Myr old (e.g., Wyatt 2003; Meyer et al.~2006).

Only when orbital migration is sufficiently smooth and slow can resonances
trap bodies. The slowness criterion requires migration to be adiabatic: 
Over the
time the planet takes to migrate across the width of the resonance, its
resonant partner must complete at least a few librations.
Otherwise the bodies speed past resonance (e.g.,
Dermott, Malhotra, \& Murray 1988; Chiang 2003; Quillen 2006).
Smoothness requires that changes in the planet's orbit which are incoherent over timescales shorter than the libration time do not accumulate unduly.
Orbital migration driven by gravitational
scattering of discrete planetesimals is intrinsically not smooth.
A longstanding concern has been whether Neptune's migration was too ``noisy''
to permit resonance capture and retention (see, e.g., Morbidelli, Brown,
\& Levison 2003). In N-body simulations of migration within planetesimal
disks (Hahn \& Malhotra 1999; Gomes et al.~2004; Tsiganis et al.~2005),
N $\sim \mathcal{O}(10^4)$ is still too small to produce the large,
order-unity capture efficiencies seemingly demanded by the current census
of Resonant KBOs.

At the same time, the impediment against resonance capture introduced
by inherent stochasticity has been exploited to
explain certain puzzling features of the Kuiper belt, most notably the
Classical (non-Resonant) belt's outer truncation radius,
assumed to lie at a heliocentric distance of $\sim$48 AU
(Trujillo \& Brown 2001; Levison \& Morbidelli 2003).
If Neptune's 2:1 resonance captured
KBOs and released them en route, Classical KBOs could have been transported
(``combed'') outwards to populate the space interior to the final position
of the 2:1 resonance, at a semi-major axis of 47.8 AU (Levison \& Morbidelli 2003).
As originally envisioned,
this scenario requires that $\sim$$3 M_\earth$ be trapped inside the 2:1
resonance so that an attendant secular resonance suppresses
growth of eccentricity during transport. It further requires that the
degree of stochasticity be such that the migration is neither too smooth
nor too noisy. Whether these requirements were actually met remain
open questions.\footnote{While
Classical KBOs do have semi-major axes that extend up to 48 AU, the
distribution of their perihelion distances cuts off sharply at
distances closer to 45 AU (see, e.g, Figure 2 of Chiang et al.~2006).
Interpreted naively (i.e., without statistics), the absence of bodies
having perihelion distances of 45--48 AU and eccentricities less
than $\sim$0.1 smacks of observational bias and motivates us to re-visit the
problem of whether an edge actually exists, or at least whether the edge
bears any relation to the 2:1 resonance.}

Stochastic migration has also been studied in gas disks, in which
noise is driven by density fluctuations in turbulent gas. Laughlin,
Steinacker, \& Adams (2004) and Nelson (2005) propose that stochasticity
arising from gas that is unstable to the magneto-rotational instability (MRI)
can significantly prolong a planet's survival time against accretion onto
the parent star. The spectrum of density fluctuations is computed
by numerical simulations of assumed turbulent gas.

In this work, we study stochastic changes to a planet's orbit
due to planetesimal scatterings. The planet's
Brownian motion arises from both Poisson variations in the rate at which
a planet encounters planetesimals, and from random fluctuations in the
mix of planet-planetesimal encounter geometries. How does the vigor of
a planet's random walk depend on the masses and orbital properties of
surrounding planetesimals? We answer this question in \S\ref{sec-oom}
by constructing an analytic theory
for how a migrating planet's semi-major axis fluctuates about its mean
value. We employ order-of-magnitude physics, verifying our assertions
whenever feasible by tailored numerical integrations. Because the properties
of planetesimal disks during the era of planetary migration are so
uncertain, we consider a wide variety of possibilities for how
planetesimal semi-major axes and eccentricities are distributed.
One of the fruits of our labors will be identification of the
conditions under which a planet's migration is maximally stochastic.

Apportioning a fixed disk mass to fewer, larger planetesimals renders
migration more noisy. How noisy is too noisy for resonance capture?
What limits can we place on the sizes of planetesimals that would
keep capture of Resonant KBOs by a migrating Neptune a viable hypothesis?
These questions are answered in \S\ref{sec-sizes}, where we write down
a simple analytic formula for the retention
efficiency of a resonance as a function of disk properties, including
planetesimal size. Quantifying the size spectrum of planetesimals
is crucial for deciphering the history of planetary systems. Many scenarios
for the evolution of the Kuiper belt implicitly assume that most of
the mass of the primordial outer solar system was locked in planetesimals
having sizes of $\mathcal{O}(100)$
km, like those observed today (see, e.g., Chiang et al.~2006 for a critique of these scenarios).
By contrast, coagulation simulations place the bulk of the mass
in bodies having sizes of $\mathcal{O}(1)$ km (Kenyon \& Luu 1999).
For ice giant formation
to proceed {\it in situ} in a timely manner in the outer solar system,
most of the primordial disk may have to reside in
small, sub-km bodies (Goldreich, Lithwick, \& Sari~2004).

In \S\ref{sec-sum}, in addition to summarizing our findings, we extend
them in a few directions.  The main thrust of this paper is to analyze
how numerous, small perturbations to a planet's orbit accumulate.
We extend our analysis in \S\ref{sec-sum} to quantify the circumstances
under which a single kick to the planet from an extremely large
planetesimal can disrupt the resonance.
We also examine perturbations exerted directly on
Resonant KBOs by ambient planetesimals.

\section{STOCHASTIC MIGRATION: AN ORDER-OF-MAGNITUDE THEORY}\label{sec-oom}  
  
We assume the planet's eccentricity is negligibly small.  We decompose the rate of change of the planet's semi-major axis, $\dot a_{\rm p}$, into average and random components,  
\begin{equation}  
\dot a_{\rm p} = \dot a_{\rm p,avg} + \dot a_{\rm p,rnd} \,\, .    
\end{equation}  
The average component (``signal'') arises from any global asymmetry in the way a planet scatters planetesimals, e.g., an asymmetry due to 
systematic differences between planetesimals inside and outside 
a planet's orbit.  The random component (``noise'') results from 
chance variations in the numbers and orbital elements of planetesimals 
interacting with the planet. By definition, $\dot a_{\rm p,rnd}$ 
time-averages to zero.  We assume that $\dot a_{\rm p,avg}(t)$ 
is a known function of time $t$, 
and devote all of \S\ref{sec-oom} to the derivation of 
$\dot a_{\rm p,rnd}$.    
  
Each close encounter between the planet and a single planetesimal lasting time $\Delta t_{\rm e}$ causes the planet's semi-major axis to change by $\Delta a_{\rm p}$. Expressions for $\Delta a_{\rm p}$ and $\Delta t_{\rm e}$ depend on the planetesimal's orbital elements. We define $x \equiv a-a_{\rm p}$ as the difference between the semi-major axes of the planetesimal and of the planet, $b > 0$ as the impact parameter of the encounter, and $u \sim e\Omega a$ as the planetesimal's random (epicyclic) velocity, where $a$, $e$, and $\Omega$ are the semi-major axis, eccentricity, and mean angular velocity of the planetesimal, respectively.  We assume that $|x| \lesssim a_{\rm p}$.  Encounters unfold differently according to how $|x|$ and $b$ compare with the planet's Hill radius,  
\begin{equation}  
R_{\rm H} = a_{\rm p}\left(\frac{M_{\rm p}}{3M_*}\right)^{1/3}  \,\, ,  
\end{equation}  
and according to how $u$ compares with the Hill velocity,  
\begin{equation}  
v_{\rm H} \equiv \Omega_{\rm p} R_{\rm H} \,\, .  
\end{equation}  
Here $M_{\rm p}$ and $M_*$ are the masses of the planet and of the star, respectively, and $\Omega_{\rm p}$ is the angular velocity of the planet.  See Table \ref{tbl-symbol} for a listing of frequently
used symbols.

In \S\ref{sec-out}, we calculate $\Delta a_{\rm p}$ for a single encounter with a planetesimal having $|x|\gtrsim R_{\rm H}$.  In \S\ref{sec-in}, we repeat the calculation for $|x|\lesssim R_{\rm H}$.  In \S\ref{sec-dom}, we provide formulae for the root-mean-squared (RMS) random velocity due to cumulative encounters, $\langle\dot a_{\rm p,rnd}^2\rangle^{1/2}$, and identify which cases of those treated in \S\S\ref{sec-out}--\ref{sec-in} potentially yield the strongest degree of stochasticity in the planet's migration.  
  
\newpage
\subsection{Single Encounters with $|x|\gtrsim R_{\rm H}$: Non-Horseshoes}\label{sec-out}   
  
We calculate the change in the planet's semi-major axis, $\Delta a_{\rm p}$, resulting from an encounter with a single planetesimal having $|x|\gtrsim R_{\rm H}$.  We treat planetesimals on orbits that do not cross that of the planet in \S\ref{sec-nonc} and those that do cross in \S\ref{sec-cross}.  Throughout, $\Delta$ refers to the change in a quantity over a single encounter, evaluated between times well before and well after the encounter.  
  
\subsubsection{Non-Crossing Orbits}\label{sec-nonc}  
Planetesimals on orbits that do not cross that of the planet have   
\begin{equation}  
|x| > ae \,\, ,  
\end{equation}  
which corresponds to  
\begin{equation}  
|x|/R_{\rm H} > u/v_{\rm H} \,\, .  
\end{equation}  
Our plan is to relate $\Delta a_{\rm p}$ to $\Delta x$ by conservation of energy, calculate $\Delta e$ using the impulse approximation, and finally generate $\Delta x$ from $\Delta e$ by conservation of the Jacobi integral.  
  
By conservation of energy,  
\begin{equation}\label{eqn-energy}  
\Delta\left[-\frac{GM_*M_{\rm p}}{2a_{\rm p}}-\frac{GM_*m}{2a}\right] = 0 \,\, ,  
\end{equation}  
where $m \ll M_{\rm p}$ is the mass of the planetesimal.  We have dropped terms that account for the potential energies of the planet and of the planetesimal in the gravitational field of the ambient disk.  These are small because the disk mass is of order $M_{\rm p} \ll M_*$ and because the disk does not act as a point mass but is spatially distributed.    
Equation (\ref{eqn-energy}) implies  
\begin{equation}\label{eqn-delap}  
\Delta a_{\rm p} \sim -\frac{m}{M_{\rm p}}\left(\frac{a_{\rm p}}{a}\right)^2 \Delta a \,\, .  
\end{equation}  
Since $|\Delta a_{\rm p}| \ll |\Delta a|$ and $a_{\rm p} \sim a$, we have $\Delta x \sim \Delta a$ and   
\begin{equation}\label{eqn-dela1}  
\Delta a_{\rm p} \sim -\frac{m}{M_{\rm p}}\Delta x \,\, .  
\end{equation}  
  
The impulse imparted by the planet changes the eccentricity of the planetesimal by $\Delta e$.  An encounter for which $|x|$ is more than a few times $R_{\rm H}$ imparts an impulse per mass\footnote{  
The impulse to the planetesimal changes both $u$ and the planetesimal's Keplerian shearing velocity, $-(3/2)\Omega x$.  In the non-crossing case, $|\Delta u| > |\Delta (\Omega x) |$.   
}  
\begin{equation}\label{eqn-impulse}  
\Delta u \sim \pm\frac{GM_{\rm p}}{b^2} \Delta t_{\rm e} \,\, .  
\end{equation}  
The impact parameter $b$ is limited by  
\begin{equation}\label{eqn-b1}  
|x|-ae \lesssim b \lesssim |x|+ae \,\, .  
\end{equation}  
Since $ae < |x|$,   
\begin{equation}  
b \sim |x| \,\, .  
\end{equation}  
Because the relative speed due to Keplerian shear, $(3/2)\Omega_{\rm p} |x|$, is larger than $u$, the relative speed during encounter is dominated by the former, and   
\begin{equation}\label{eqn-delt1}  
\Delta t_{\rm e} \sim \frac{2b}{(3/2)\Omega_{\rm p} b} =\frac{4}{3\Omega_{\rm p}}\sim \frac{1}{\Omega_{\rm p}} \,\, .  
\end{equation}  
Since $\Delta t_{\rm e}$ is about one-fifth of an orbital period, the impulse approximation embodied in (\ref{eqn-impulse}) should yield good order-of-magnitude results.  The change in the eccentricity of the planetesimal is hence  
\begin{equation}\label{eqn-dele1}  
\Delta e \sim \frac{\Delta u}{\Omega_{\rm p} a_{\rm p}} \sim \pm\frac{M_{\rm p}}{M_*}\left(\frac{a_{\rm p}}{x}\right)^2 \,\, .  
\end{equation}  
When $|\Delta e| <$ (the pre-encounter) $e$,  
the change $\Delta e$ can be either positive or negative, 
depending on the true anomaly of the planetesimal at the time of encounter. 
If $|\Delta e| > e$, then $\Delta e > 0$.  
When $|x|\sim R_{\rm H}$, $|\Delta e|$ attains its maximum value of $\sim$$(M_{\rm p}/M_*)^{1/3}$; i.e., $\Delta u \sim v_{\rm H}$.\footnote{  
When $|x|\lesssim 2R_{\rm H}$, the encounter pulls the planetesimal into the planet's Hill sphere.    
The planetesimal accelerates in a complicated way and exits the Hill sphere in a random direction with $u$ of order the planet's escape velocity at the Hill radius, $v_{\rm H}$ (Petit \& H\'enon 1986).  The planetesimal's eccentricity is boosted by $\Delta e \sim (M_{\rm p}/M_*)^{1/3}$.  
The encounter time is typically the time required to complete a few orbits around the planet, $\Delta t_{\rm e}\sim 2\pi/\Omega_{\rm p}$.  
Since $\Delta t_{\rm e}$ and $\Delta e$ match, to order of magnitude, Equations (\ref{eqn-delt1}) and (\ref{eqn-dele1}) for $|x|\sim R_{\rm H}$, we do not treat $R_{\rm H} \lesssim |x|\lesssim 2R_{\rm H}$ as an explicitly different case.  
}    
  
To calculate the corresponding change in the planetesimal's semi-major axis, $\Delta x$, we exploit conservation of the Jacobi integral, $C_{\rm J}$.  That a conserved integral exists relies on the assumption that in the frame rotating with the planet, the potential (having a centrifugal term plus gravitational contributions due to the star, planet, and disk) is time-stationary; the Jacobi integral is simply the energy of the planetesimal (test particle) evaluated in that frame. To the same approximation embodied in Equation (\ref{eqn-energy}), 
\begin{eqnarray} \label{eqn-cj} 
-\frac{1}{2} C_{\rm J} &=& E -\Omega_{\rm p} J \nonumber \\
&=& -\frac{GM_*}{a_{\rm p}}\left[\frac{1}{2(a/a_{\rm p})} + \sqrt{(a/a_{\rm p})(1-e^2)} \right] 
\end{eqnarray}  
far from encounter, where $E$ and $J$ are the energy and angular momentum per mass of the planetesimal, respectively. Taking the differential of (\ref{eqn-cj}) yields, to leading order,  
\begin{equation}\label{eqn-jac}  
\frac{3}{4}\left(\frac{\Delta x}{a_{\rm p}}\right)^2 + \left(\frac{3}{2}\frac{x}{a_{\rm p}} - \frac{1}{2}e^2\right)\frac{\Delta x}{a_{\rm p}} -\Delta(e^2) = 0 \,\, .  
\end{equation}  
Since $|x|/a_{\rm p} > e > e^2$ (non-crossing condition) and $\Delta(e^2) < (x/a_{\rm p})^2$ (by Equation [\ref{eqn-dele1}] and the condition $|x|>R_{\rm H}$), Equation (\ref{eqn-jac}) reduces to  
\begin{equation}\label{eqn-delx1}  
\Delta x \sim \frac{2a_{\rm p}^2}{3x}\Delta (e^2) \,\, .  
\end{equation}  
We combine Equations (\ref{eqn-dela1}) and (\ref{eqn-delx1}) to find  
\begin{equation}\label{eqn-dela2}  
\Delta a_{\rm p} \sim -\frac{m}{M_{\rm p}}\frac{a_{\rm p}^2}{x}\Delta (e^2) \,\, .  
\end{equation}  
  
Equation (\ref{eqn-dela2}) takes two forms depending on how $|\Delta e|$ compares with (the pre-encounter) $e$.  If   
\begin{equation}  
|x| > R_{\rm H} \left(\frac{v_{\rm H}}{u}\right)^{1/2} \,\, ,  
\end{equation}  
then $|\Delta e| < e$, $\Delta (e^2) \sim 2e\Delta e$, and Equation (\ref{eqn-dela2}) becomes  
\begin{equation}\label{eqn-dela3}  
\Delta a_{\rm p} \sim \mp\frac{m}{M_*}\frac{a_{\rm p}^4}{x^3}e \,\, .  
\end{equation}  
The right-hand side is extremized for $|x|\sim R_{\rm H}(v_{\rm H}/u)^{1/2}$:  
\begin{equation}\label{eqn-max1}  
\max|\Delta a_{\rm p}| \sim \frac{m}{M_{\rm p}}R_{\rm H}\left(\frac{a_{\rm p}e}{R_{\rm H}}\right)^{5/2} <  \frac{m}{M_{\rm p}}R_{\rm H} \,\, ,  
\end{equation}  
valid for $e<R_{\rm H}/a_{\rm p}$ (non-crossing).  
  
On the other hand, if   
\begin{equation}  
R_{\rm H} \lesssim\, x < R_{\rm H} \left(\frac{v_{\rm H}}{u}\right)^{1/2} \,\, ,  
\end{equation}  
then $|\Delta e| > e$ and $\Delta (e^2) \sim (\Delta e)^2$, and Equation (\ref{eqn-dela2}) becomes  
\begin{equation}\label{eqn-dela4}  
\Delta a_{\rm p} \sim -\frac{mM_{\rm p}}{M_*^2}\frac{a_{\rm p}^6}{x^5} \,\, .  
\end{equation}  
Equation (\ref{eqn-dela4}) agrees with the more careful solution of Hill's problem by H\'enon \& Petit (1986). The right-hand side is extremized for $|x|\sim R_{\rm H}$:   
\begin{equation}\label{eqn-max2}  
\max|\Delta a_{\rm p}| \sim \frac{m}{M_{\rm p}}R_{\rm H} \,\, .  
\end{equation}

In summary, if $x/R_{\rm H} > u/v_{\rm H}$, then (a) the planetesimal's orbit does not cross (``nc'') that of the planet, (b)  
\begin{equation}\label{eqn-delt}  
\Delta t_{\rm e}  \sim \frac{1}{\Omega_{\rm p}} \,\, ,  
\end{equation}  
and (c)  
\begin{equation}\label{eqn-sub}  
\Delta a_{\rm p} =  
   \left\{\begin{array}{l}
     \Delta a_{\rm p,nc1} \sim  -{\displaystyle\frac{mM_{\rm p}}{M_*^2}\frac{a_{\rm p}^6}{x^5}}, \\\qquad\qquad\qquad\qquad\mbox{if $R_{\rm H} \lesssim x  \lesssim  (v_{\rm H}/u)^{1/2}R_{\rm H}$;} \\  
    \rule{0ex}{5ex}\Delta a_{\rm p,nc2}  \sim \mp {\displaystyle\frac{m}{M_*}\frac{a_{\rm p}^4}{x^3}}e, \\\qquad\qquad\qquad\qquad\mbox{if $x  \gtrsim   (v_{\rm H}/u)^{1/2}R_{\rm H}$.}   
  \end{array}\right.  
\end{equation}

\subsubsection{Crossing Orbits}\label{sec-cross}  
  
Encounters with planetesimals on orbits that cross that of the planet, i.e., 
those with   
\begin{equation}  
|x|/R_{\rm H} < u/v_{\rm H} \,\, ,  
\end{equation}  
differ from encounters with non-crossing planetesimals in two key respects.  First, the relative velocity of the two bodies is dominated by the planetesimal's random (epicyclic) velocity rather than the Keplerian shear. Second, the planetesimal's impact parameter, $b$, may differ significantly from $|x|$.  The impact parameter may take any value  
\begin{equation}  
b_{\rm min} < b \lesssim ae \,\, ,  
\end{equation}  
where $b_{\rm min}$ is the impact parameter below which the planetesimal collides with the planet. Because crossing orbits allow for encounters with many different geometries, outcomes of these encounters can vary dramatically.  Here we restrict ourselves to estimating the maximum $|\Delta a_{\rm p}|$ that can result from an orbit-crossing encounter. In \S\ref{sec-super}, 
we argue this restriction is sufficient for our purposes.  
  
When $u > v_{\rm H}$, the eccentricity of the planetesimal can change by at most $|\Delta e| \sim e$.  Such a change corresponds to an order-unity rotation of the direction of the planetesimal's random velocity vector, and requires that $b \lesssim GM_{\rm p}/u^2$.  The change in the planetesimal's specific energy over the encounter is approximately   
\begin{equation} \label{eqn-delenergy}  
\Delta\left(-\frac{GM_*}{2a}\right) \sim \Delta\left(\frac{1}{2}v^2\right) + \Delta\left(-\frac{GM_*}{r}\right) \,\, ,  
\end{equation}  
where $v$ is the velocity of the planetesimal relative to the star (in an inertial frame of reference) and $r$ is the distance between the planetesimal and the star.  Now $\Delta r$ for an encounter with $b \lesssim GM_{\rm p}/u^2$ is at most $GM_{\rm p}/u^2<R_{\rm H}$ and $\Delta(v^2)$ is of order $\Omega au$.  Since  
\begin{equation}  
\Omega au \gtrsim \Omega^2 a R_{\rm H} > \Omega^2 a\frac{GM_{\rm p}}{u^2} \,\, ,  
\end{equation}  
the second term on the right-hand side of Equation (\ref{eqn-delenergy}) is negligible compared to the first, and the maximum $|\Delta a|$ over an encounter is 
\begin{equation}\label{eqn-ae}  
\max|\Delta a| \sim \frac{a^2}{GM_*}\Omega a u \sim ae \,\, .  
\end{equation}  
By Equation (\ref{eqn-delap}) and $a\sim a_{\rm p}$,  
\begin{equation}\label{eqn-delamax}  
\max\left| \Delta a_{\rm p}\right| \sim \frac{m}{M_{\rm p}}a_{\rm p}e \,\, .  
\end{equation}  
We have verified Equation (\ref{eqn-delamax}) by numerical orbit 
integrations. We could also have arrived at Equation (\ref{eqn-delamax}) 
through Equation (\ref{eqn-jac}), which yields $|\Delta x| \sim a_{\rm p} e$ 
for crossing orbits when $|\Delta e|\sim e$.

\subsection{Single Encounters with $|x|\lesssim R_{\rm H}$: Horseshoes}\label{sec-in}  
  
When $|x| < R_{\rm H}$, planetesimals can occupy horseshoe orbits.  A planetesimal on a horseshoe orbit for which $|x| \approx R_{\rm H}$ encounters the planet on a timescale somewhat shorter than the orbital period; by the impulse approximation, such a planetesimal kicks the planet such that  
\begin{equation}  
\Delta a_{\rm p} \sim \frac{m}{M_*}\frac{a_{\rm p}^3}{b^2} \,\, ,  
\end{equation}  
where 
we have momentarily restricted consideration to planetesimals 
having sub-Hill eccentricities ($e \lesssim R_{\rm H}/a_{\rm p}$).  
From H\'enon and Petit (1986),  
\begin{equation}  
b = \frac{8}{3}\frac{R_{\rm H}^3}{x^2},  
\end{equation}  
valid for $x$ not too far below $R_{\rm H}$.  
Then  
\begin{equation}  
|\Delta a_{\rm p}| \sim \frac{m}{M_{\rm p}}\frac{x^4}{R_{\rm H}^3} \,\, .  
\end{equation}  
The kick is maximal for maximum $|x| = R_{\rm H}$:  
\begin{equation}  
\max|\Delta a_{\rm p}| \sim \frac{m}{M_{\rm p}}R_{\rm H} \,\, .  
\end{equation}  
This is the same maximum as was derived for the $|x| \sim R_{\rm H}$, non-crossing case; see Equation 
(\ref{eqn-max2}).  Thus, a co-orbital ring of planetesimals on horseshoe orbits with sub-Hill eccentricities increases the stochasticity generated by planetesimals on non-horseshoe, non-crossing orbits by a factor of at most order unity (under the assumption that disk properties are roughly constant within several Hill radii of the planet).  For this reason, and also because the horseshoe region may well have been depleted of planetesimals compared to the rest of the disk, we omit consideration of co-orbital, sub-Hill planetesimals for the remainder of the paper, confident that the error so incurred will be at most order unity.  
  
What about planetesimals on horseshoe orbits with super-Hill eccentricities ($e \gtrsim R_{\rm H}/a_{\rm p}$)?  Upon encountering the planet, such objects can have their semi-major axes changed by $|\Delta a| > R_{\rm H}$---whereupon they are expelled from the 1:1 horseshoe resonance.  Because highly eccentric, horseshoe resonators are unstable, we neglect consideration of them for the rest of our study.  
   
\subsection{Multiple Encounters: Cumulative Stochasticity}\label{sec-dom}  
  
We now extend our analysis from individual encounters to the cumulative stochasticity generated by a disk with surface density 
$\Sigma_m$ in planetesimals of a single mass $m$.
Note that $\Sigma_m$ need not equal the total surface density $\Sigma$
(integrated over all possible masses $m$). We will consider
size distributions in \S\ref{sec-sizedist}.
We consider planetesimals with sub-Hill ($u < v_{\rm H}$) velocities (\S\ref{sec-sub}) separately from those with super-Hill ($u > v_{\rm H}$) velocities (\S\ref{sec-super}). Sub-Hill (non-horseshoe) planetesimals always occupy non-crossing orbits. Super-Hill planetesimals can be crossing or non-crossing.

\subsubsection{Sub-Hill Velocities ($u<v_{\rm H}$)}\label{sec-sub}

Consider planetesimals with sub-Hill velocities located  
a radial distance $x$ away from the planet ($|x| > R_{\rm H}$).  Since $u < v_{\rm H}$, the speeds of planetesimals relative to the planet are 
determined principally by Keplerian shear (Equation [\ref{eqn-delt1}]), and the scale height of the planetesimals is less than $R_{\rm H}$.  
The planet encounters (undergoes conjunctions with) such  
planetesimals at a mean rate  
\begin{equation}\label{eqn-subencounter}  
\dot{\overline{N}} \sim \frac{\Sigma_m}{m} \Omega x^2 \,\,,  
\end{equation}  
as is appropriate for encounters in a two-dimensional geometry.  
Over a time interval $\Delta t$, the planet encounters  
$\overline{N} = \dot{\overline{N}}\Delta t$ such planetesimals on average.  
Systematic trends in $\overline{N}$ with $x$---say, systematically more  
objects encountered interior to the planet's orbit than exterior  
to it---cause the planet to migrate along an average trajectory with  
velocity $\dot{a}_{\rm p,avg}$.

Random fluctuations in (a) the number
of planetesimals encountered per fixed time interval and (b) the mix of planetesimals' pre-encounter
orbital elements cause the planet to random walk about
this average trajectory. Contribution (a) is straightforward to model.
The probability that the planet encounters
$N$ objects located a distance $x$ away in time $\Delta t$ is given by Poisson
statistics:
\begin{equation}\label{eqn-poiss}
P(N) = \frac{{\overline{N}}^N}{N!}e^{-\overline{N}} \,\, .
\end{equation}
The variance in $N$ is
\begin{equation}
\sigma_N^2 \equiv \left<(N-\overline N)^2\right> = \overline N \,\,.
\end{equation}
Fluctuations in $N$ drive the planet either towards or away
from the star with equal probability and with typical speed
\begin{equation}
\langle \dot{a}_{\rm p,rnd}^2 \rangle^{1/2} \sim \frac{ |\Delta a_{\rm
p}|}{\Delta t}\overline{N}^{1/2} \,,
\label{eqn-general}
\end{equation}
hereafter the root-mean-squared (RMS) speed.
While $\langle \dot{a}_{\rm p,rnd}^2 \rangle^{1/2} \propto 1/\sqrt{\Delta t}$,
the distance random walked $\langle \dot{a}_{\rm p,rnd}^2
\rangle^{1/2} \Delta t \propto \sqrt{\Delta t}$.

Our assumption of Poisson statistics is reasonable. In the
sub-Hill case, a planet-planetesimal encounter requires a time
$\Delta t_{\rm e} \sim 1/\Omega_{\rm p}$ (Equation [\ref{eqn-delt1}]) to complete.
Encounters separated by more than $\Delta t_{\rm e}$ are uncorrelated
with one another, at least until the planet completes one revolution
with respect to the surrounding disk, i.e., at least until a synodic
time $t_{\rm syn} \sim 4\pi a_{\rm p}/(3\Omega_{\rm p} |x|)$ elapses. After a synodic
period, it is possible, in principle, for the planet to essentially
repeat the same sequence of encounters that it underwent during the
last synodic period. We assume in this paper that this does not
happen---that the orbits of planetesimals interacting with the planet
are randomized on a timescale $t_{\rm rdz} < t_{\rm syn}$.
We expect this inequality to be enforced by a combination of
(i) randomization of planetesimal orbits due to encounters
with the planet (e.g., encounters within the chaotic zone
of the planet [Wisdom 1980]), (ii) phase mixing of planetesimals
due to Keplerian shear (which occurs on timescale $t_{\rm syn}$
for planetesimals distributed between $x$ and $\sim$$2x$),
(iii) gravitational interactions between planetesimals,
and (iv) physical collisions between planetesimals.
As long as $t_{\rm rdz} < t_{\rm syn}$, we are free to choose
$\Delta t$ to be anything longer than $\Delta t_{\rm e}$.\footnote{
If $t_{\rm rdz} > t_{\rm syn}$, then
$\Delta t > t_{\rm rdz}$ and the right-hand
side of Equation (\ref{eqn-general}) is multiplied
by $\sqrt{t_{\rm rdz}/t_{\rm syn}}$. The planet's motion is 
more stochastic in this case because
over $t_{\rm rdz}$, correlated interactions with planetesimals
do not cancel each other as much as uncorrelated interactions
would.  Later, since we will be interested in stochastic perturbations 
to mean-motion resonant particles, we will require $t_{\rm rdz} < t_{\rm lib}$, where $t_{\rm lib}$ is the libration period within resonance.}

Contribution (b) is difficult to model precisely since we do not know
how orbital elements of planetesimals are distributed. These
distributions are unlikely to be governed by simple Poisson or Gaussian
statistics (see, e.g., Ida \& Makino 1992; Rafikov 2003; Collins \& Sari 2006).  
Nonetheless,
neglecting contribution (b) will not lead to serious error. Suppose the
planetesimals' pre-encounter elements are distributed such that the
fractional variation in each element is at most of order unity
(e.g., the planetesimal eccentricities span a range from $e/2$ to $2e$
at most).
Then the central limit theorem ensures that the noise introduced
by random sampling of orbital elements is at most comparable to the noise
introduced by random fluctuations in the encounter rate.
Consider, for example, noise that arises from random sampling of $e$
in the case where $\Delta a_{\rm p} = \Delta a_{\rm p,nc2}$
(Equation [\ref{eqn-sub}]). For an encounter rate fixed at
$\dot{\overline N}$, the planet's semi-major axis
$a_{\rm p}$ changes over time interval
$\Delta t$ by ${\overline N} \times \overline{\Delta a_{\rm p}}$,
where $\overline{\Delta a_{\rm p}}$ is the mean of
$\overline{N} = \dot{\overline N} \Delta t$ sampled values
of $\Delta a_{\rm p}$.
If the dispersion in $e$ for individual planetesimals
is $\sigma_e$ and the mean eccentricity sampled over ${\overline N}$
values is $\overline{e}$,
then the dispersion in the sampled mean eccentricity is
$\sigma_{\overline{e}} \sim \sigma_e / \overline{N}^{1/2}$
by the central limit theorem.
For $\Delta a_{\rm p} = \Delta a_{\rm p,nc2} \propto e$,
the dispersion in $\overline{\Delta a_{\rm p}}$ is
$|\Delta a_{\rm p}| \sigma_{\overline{e}}/e$.
The planet's RMS speed generated purely from random sampling of $e$ is
\begin{equation}
\langle \dot{a}_{\rm p,rnd}^2 \rangle^{1/2} \sim
\frac{\overline{N}}{\Delta t}\, \left| \Delta a_{\rm p} \right| \,
\frac{\sigma_{\overline{e}}}{e} \sim \frac{|\Delta a_{\rm p}|}{\Delta
t}\, \overline{N}^{1/2} \, \frac{\sigma_e}{e} \,\, ,
\end{equation}
which is at most comparable to the RMS speed generated purely from
random sampling of $N$ (Equation [\ref{eqn-general}]), as desired.
Our original supposition
that $\max(\sigma_e/e) \sim 1$ leads to no important loss of generality;
if the distribution in $e$ were bi-modal, for example, we could
treat each population separately and add the resultant RMS speeds in
quadrature. Similar results obtain for random sampling of other
elements such as $x$.
For simplicity, we hereafter treat explicitly only fluctuations
in the encounter rate [(Equation (\ref{eqn-general})],
knowing that the noise
so calculated will be underestimated by a factor of at most order unity.

Expression (\ref{eqn-general})  
measures the contribution to the RMS speed from planetesimals located  
a distance $x$ away. Since $\Delta a_{\rm p}$ scales inversely with $x$  
to a steep power in the sub-Hill regime (see [\ref{eqn-sub}]),  
the contribution to the RMS speed is greatest from objects at small $x$  
(for reasonable variations of $\Sigma_m$ with $x$).  
We take disk material to extend to a minimum distance of 
$|x_{\rm min}| \equiv \mathcal{R}R_{\rm H}$ ($\mathcal{R} > 1$)  
from the planet's orbit. 
Insertion of (\ref{eqn-sub}) into (\ref{eqn-general}) yields  
\begin{equation}\label{eqn-subnoise-pre}  
\left<\dot a_{\rm p,rnd}^2\right>^{1/2} \sim  
  \left\{\begin{array}{l}  
     {\displaystyle\mathcal{R}^{-4} \frac{1}{(\Omega_{\rm p}\Delta t)^{1/2}} \left( \frac{\Sigma_m a_{\rm p}^2 m}{M_{\rm p}^2} \right)^{1/2} \frac{R_{\rm H}}{a_{\rm p}} v_{\rm H}}, \\ \qquad\qquad\qquad\qquad\rule{0in}{0.15in}\mbox{if $1 < \mathcal{R} < (v_{\rm H}/u)^{1/2}$;} \\  \\
      \rule{0ex}{5ex}{\displaystyle\mathcal{R}^{-2} \frac{1}{(\Omega_{\rm p}\Delta t)^{1/2}} \left( \frac{\Sigma_m a_{\rm p}^2 m}{M_{\rm p}^2} \right)^{1/2} e v_{\rm H}}, \\ \qquad\qquad\qquad\qquad\rule{0in}{0.15in}\mbox{if $\mathcal{R} > (v_{\rm H}/u)^{1/2}$.}   
  \end{array}\right.  
\end{equation}  
\noindent The RMS speed is maximized for $\mathcal{R} = 1$.  
 
\subsubsection{Super-Hill Velocities ($u>v_{\rm H}$)}\label{sec-super}  

Next we consider the noise generated by planetesimals with  
super-Hill random velocities ($u>v_{\rm H}$). We refer to 
close encounters that change 
$a_{\rm p}$ by the maximal 
amount, 
$\max|\Delta a_{\rm p}| \sim (m/M_{\rm p}) a_{\rm p}e$ 
(Equation [\ref{eqn-delamax}]), as ``maximal encounters.'' 
Maximal encounters, which occur at impact parameters
$b \lesssim GM_{\rm p}/u^2$, make an order-unity contribution to 
the total super-Hill stochasticity.
Non-maximal (more distant) encounters contribute
to the total stochasticity through a Coulomb-like logarithm,
as we show at the end of this sub-section.

For a maximal encounter, 
a planetesimal must approach within distance $b \lesssim GM_{\rm p}/u^2$.  
Such encounters occur at a mean rate   
\begin{equation}\label{eqn-Ndot}  
\dot{\overline N} \sim n\left(\frac{GM_{\rm p}}{u^2}\right)^2 u \sim \frac{\Sigma_m}{m}\Omega_{\rm p}R_{\rm H}^2\left(\frac{v_{\rm H}}{u}\right)^4 \,\, ,  
\end{equation}  
where $n\sim\Sigma_m\Omega/(mu)$ is the number density of planetesimals, and we have assumed that planetesimal inclinations and eccentricities are of the same order.  Over a time interval $\Delta t$, the planet encounters on average $\dot{\overline N}\Delta t$ such planetesimals, each of which increases or decreases $a_{\rm p}$ by about $\max|\Delta a_{\rm p}|$.  Since planetesimals suffering maximal encounters have their orbits effectively randomized relative to each other, we may choose $\Delta t$ to be any time interval longer than an encounter time  $\Delta t_{\rm e} \sim b/u < 1/\Omega_{\rm p}$ (see related discussion in \S\ref{sec-sub}).  Therefore the planet random walks with RMS velocity (averaged over time $\Delta t$)  
\begin{eqnarray}\label{eqn-superrms}  
&&\left<\dot a_{\rm p,rnd}^2\right>^{1/2} \sim \frac{(\dot{\overline N}\Delta t)^{1/2} \max|\Delta a_{\rm p}|}{\Delta t} \nonumber \\
&&\sim \left(\frac{1}{\Omega_{\rm p}\Delta t}\right)^{1/2}\left(\frac{\Sigma_m a_{\rm p}^2 m}{M_{\rm p}^2}\right)^{1/2}\frac{v_{\rm H}}{u}\frac{R_{\rm H}}{a_{\rm p}}v_{\rm H} \,\, .  
\end{eqnarray}  

What about the contribution from
non-maximal encounters? For a super-Hill encounter at
impact parameter $b$, the specific impulse imparted to the planetesimal is
$\sim$$GM_{\rm p}/(bu)$. We suppose that
$|\Delta a_{\rm p}|$ is proportional to this specific impulse, so that
$|\Delta a_{\rm p}| \propto 1/b$. We have confirmed this
last proportionality by numerical orbit integrations (not shown).
Since $\left<\dot a_{\rm p,rnd}^2\right>^{1/2} \propto
(\dot{\overline N})^{1/2} |\Delta a_{\rm p}|$ and
$\dot{\overline N} \propto b^2$, we have
$\left<\dot a_{\rm p,rnd}^2\right>^{1/2} \propto b^0$,
which implies that each octave in impact parameter contributes
equally to the total stochasticity. In other words,
our estimate for $\left<\dot a_{\rm p,rnd}^2\right>^{1/2}$ in
Equation (\ref{eqn-superrms}) should
be enhanced by a logarithmic factor of $\ln(b_{\rm max}/b_{\rm min})$,
where $b_{\rm max}$ and $b_{\rm min} \sim GM_{\rm p}/u^2$
are maximum and minimum
impact parameters. We estimate $b_{\rm max} \sim u/\Omega$, the
value for which the relative velocity of a super-Hill encounter
is dominated by the planetesimal's random velocity rather than by
the background shear. The logarithm is not large; for example,
for $e=0.2$, $\ln[(u/\Omega)/(GM_{\rm p}/u^2)] \sim 5$.

\subsubsection{Summary} \label{sec-neatsum} 
  
We can neatly summarize Equations (\ref{eqn-subnoise-pre}) and (\ref{eqn-superrms})  
by defining the Hill eccentricity,  
\begin{equation}  
e_{\rm H} \equiv R_{\rm H}/a_{\rm p} \,\, ,  
\end{equation}  
and parameterizing $\Sigma_m$ such that the disk contains  
mass $\mathcal{M} M_{\rm p}$ in planetesimals of mass $m$ spread uniformly from $a_{\rm d}/2$  
to $3a_{\rm d}/2$:  
\begin{equation}  
\Sigma_m = \frac{\mathcal{M}M_{\rm p}}{2\pi a_{\rm d}^2} \,,  
\end{equation}  
where $\mathcal{M}$ is a dimensionless number of order unity. Then  
\begin{equation}\label{eqn-rosetta}  
\left<\dot a_{\rm p,rnd}^2\right>^{1/2} \sim  
  \left\{\begin{array}{l}  
     \displaystyle{\mathcal{C}\mathcal{R}^{-4}\left(\Omega_{\rm p}\Delta t\right)^{-1/2} \left( \frac{\mathcal{M} m}{M_{\rm p}} \right)^{1/2} \frac{a_{\rm p}}{a_{\rm d}}\,e_{\rm H}\,v_{\rm H}}, \\\qquad\qquad\qquad\qquad\rule{0in}{0.15in}\mbox{if $e<e_{\rm H}/\mathcal{R}^2$;} \\ \\ 
     \rule{0ex}{5ex}\displaystyle{\mathcal{C}\mathcal{R}^{-2}\left(\Omega_{\rm p}\Delta t\right)^{-1/2} \left( \frac{\mathcal{M} m}{M_{\rm p}} \right)^{1/2}\frac{a_{\rm p}}{a_{\rm d}} \,e\,v_{\rm H}}, \\ \qquad\qquad\qquad\qquad\rule{0in}{0.15in}\mbox{if $e_{\rm H}/\mathcal{R}^2 < e < \mathcal{R}e_{\rm H}$;} \\  \\
     \rule{0ex}{5ex}\displaystyle{\mathcal{C}\left(\Omega_{\rm p}\Delta t\right)^{-1/2}\left( \frac{\mathcal{M} m}{M_{\rm p}} \right)^{1/2}\frac{a_{\rm p}}{a_{\rm d}} \, \frac{e_{\rm H}^2}{e} \, v_{\rm H}}, \\\qquad\qquad\qquad\qquad\rule{0in}{0.15in}\mbox{if $e > \mathcal{R}e_{\rm H}$,}  
  \end{array}\right.  
\end{equation}  
where we have introduced a constant coefficient $\mathcal{C}$ 
(the same for each case so that the function remains continuous across 
case boundaries). The coefficient $\mathcal{C}$ encapsulates all the factors 
of order unity that we have dropped in our derivations. By studying 
N-body simulation data pertaining to the case $e > \mathcal{R}e_{\rm H}$ 
as recorded in the literature (Hahn \& Malhotra 1999; 
Gomes et al.~2004), we estimate that $\mathcal{C}$ is possibly of the order 
of several. That $\mathcal{C} > 1$ (but not $\gg 1$) is consonant with 
our having consistently underestimated the noise by neglecting (a) distant, 
non-maximal encounters in the super-Hill regime (\S\ref{sec-super}), and 
(b) stochasticity introduced by random sampling of orbital elements of 
planetesimals encountering the planet (\S\ref{sec-sub}). 
We defer definitive calibration of $\mathcal{C}$ to future study, but 
retain the coefficient 
in our expressions below to assess the degree to which our quantitative 
estimates are uncertain. 

Planetesimals with $e=\mathcal{R}e_{\rm H}$ ($|x_{\rm min}|=a_{\rm p}e$) produce the maximum possible stochasticity: 
\begin{eqnarray}\label{eqn-maxstoch}  
\max \left<\dot a_{\rm p,rnd}^2\right>^{1/2}   
&\sim& \frac{\mathcal{C}}{\mathcal{R}}\left(\Omega_{\rm p}\Delta t\right)^{-1/2}\left( \frac{\mathcal{M} m}{M_{\rm p}} \right)^{1/2}  \frac{a_{\rm p}}{a_{\rm d}} \, e_{\rm H} \, v_{\rm H},\nonumber \\
&& \qquad \mbox{for } e=\mathcal{R}e_{\rm H} \geq e_{\rm H} \,\, .  
\end{eqnarray}
We will often adopt this case for illustration purposes below.  

Since
$\left<\dot a_{\rm p,rnd}^2\right>^{1/2} \propto (\mathcal{M}m)^{1/2}$,
the stochasticity driven by planetesimals having a range of sizes is dominated
by those objects (in some logarithmic size bin) having maximal $\Sigma_mm$.
For common power-law size distributions, such objects will occupy
the upper end of the distribution. We will explicitly consider
various possible size distributions in \S\ref{sec-sizedist}.

\section{APPLICATION: MAXIMUM PLANETESIMAL SIZES}\label{sec-sizes}  
  
Neptune is thought to have migrated outward by scattering planetesimals 
during the late stages of planet formation (Fern\'andez \& Ip 1984; 
Hahn \& Malhotra 1999).   
As the planet migrated, it may have captured Kuiper belt objects into its  
exterior resonances (Malhotra 1995), giving rise to the Resonant KBOs 
observed today (Chiang et al.~2003; Hahn \& Malhotra 2005).  If Neptune's 
migration had been too stochastic, however,  
resonance capture could not have occurred.  
A planetesimal disk having fixed surface mass density $\Sigma$ 
generates more stochasticity when composed of larger (fewer) planetesimals. 
Therefore, assuming that planetary migration and concomitant resonance capture 
correctly explain the origin of present-day Resonant KBOs,   
we can rule out size distributions that are too ``top-heavy'' during the era of migration.
  
Stochasticity causes a planet to migrate both outward and inward.  
In \S\ref{sec-outin}, we provide background information regarding 
how resonance capture and retention depend on the sign of migration. 
In \S\ref{sec-gencon}, we lay out general considerations for whether 
a stochastically migrating planet can retain particles in resonance. 
In \S\ref{sec-oomsize},  
we derive and evaluate analytic, order-of-magnitude  
expressions for the maximum planetesimal size compatible  
with resonance retention, in the simple case when all planetesimals
have the same size. In \S\ref{sec-analyt}, we provide an analytic 
formula that details precisely how the resonance retention efficiency 
varies with average migration speed and planetesimal size. 
These analytic results are quantitatively tested  
by numerical integrations in \S\ref{sec-num}. 
Cases where planetesimals exhibit a wide range of sizes
are examined in \S\ref{sec-sizedist}.

\subsection{Migrating Outward and Inward}\label{sec-outin}  
  
As a planet migrates smoothly outward (away from the parent star), it can capture 
planetesimals into its exterior mean-motion resonances. By contrast,  
a planet which migrates smoothly inward cannot capture 
planetesimals which are initially non-resonant into its exterior resonances 
(e.g., Peale 1986). But a planetesimal that starts in exterior resonance 
with an inwardly migrating planet can remain in resonance for a finite time.  
  
Goldreich (1965) demonstrates how an outwardly migrating body adds angular momentum to a test particle in exterior resonance at just the right rate to keep the particle in resonance.  Reversing the signs in his proof implies that an inwardly migrating body removes angular momentum from a particle in exterior resonance, pulling it inward while preserving the resonant lock.    
A planetesimal's eccentricity decreases as it is pulled inward.  The adiabatic invariant,  
\begin{equation}\label{eqn-Nadiabat}  
\mathcal{N} = \sqrt{GM_*a}(p\sqrt{1-e^2}-q) \,\, ,  
\end{equation}  
which is preserved for migration timescales long compared to the synodic time (e.g., Murray-Clay \& Chiang 2005), implies that a planetesimal in $p:q$ exterior resonance ($p>q$) cannot be pulled inward to a semi-major axis less than   
\begin{equation}  
a_{\rm min} = \frac{1}{GM_*}\left(\frac{\mathcal{N}}{p-q}\right)^2 = a_0\left(\frac{p\sqrt{1-e_0^2} -q}{p-q}\right)^2 \,\, ,  
\end{equation}  
the value for which $e=0$.  Here $a_0$ and $e_0$ are the initial semi-major axis and eccentricity of the planetesimal, respectively.    
  
\begin{figure}  
\epsscale{1.2}  
\plotone{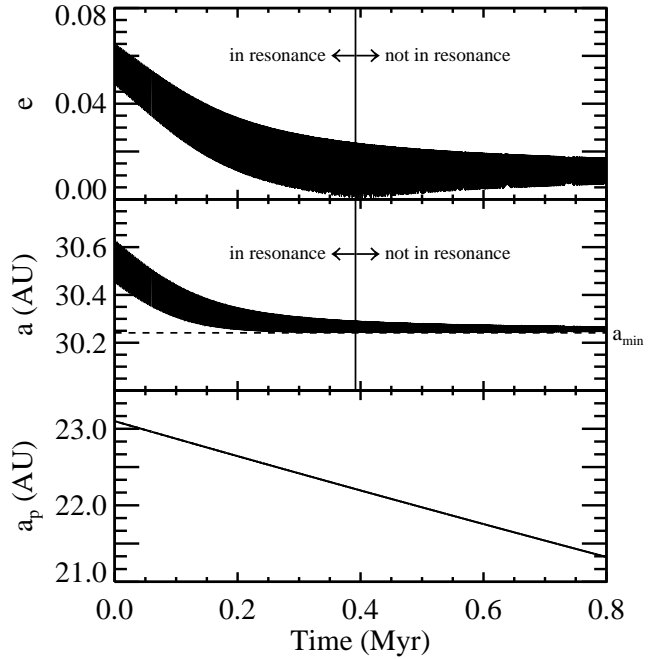}  
\caption{Evolution of a planetesimal in external 3:2 resonance as the planet migrates smoothly inward.  The planetesimal remains in resonance until its eccentricity reaches zero, at a semi-major axis of $a=a_{\rm min}$.}  
\label{fig-zeroe}  
\end{figure}

\begin{figure}  
\epsscale{1.2}  
\plotone{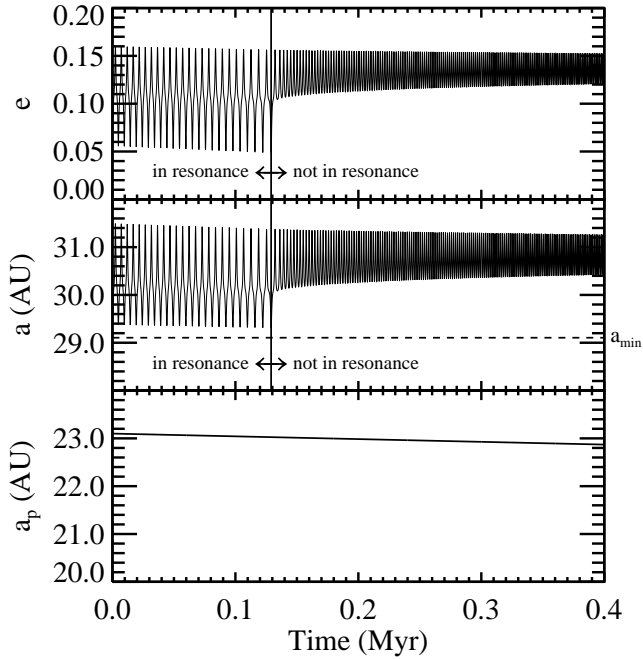}  
\caption{Evolution of a planetesimal in external 3:2 resonance as the planet migrates smoothly inward.  The planetesimal remains in resonance until it crosses the separatrix; by contrast to the evolution shown in Figure \ref{fig-zeroe}, 
the particle's semi-major axis does not reach $a_{\rm min}$ and its eccentricity stays greater than zero.}  
\label{fig-separ}  
\end{figure}

Thus an exterior particle follows an inwardly migrating planet in resonant lockstep until it either reaches zero eccentricity (view Figure 4 of Peale 1986 or Figure 8.22 of Murray \& Dermott 1999 in reverse) or until it crosses the separatrix (view Figure 5 of Peale 1986 or Figure 8.23 of Murray \& Dermott 1999 in reverse), whichever comes first.  We illustrate the former possibility in Figure \ref{fig-zeroe} and the latter possibility in Figure \ref{fig-separ}, using our own orbit integrations. The value of $a_{\rm min}$ is annotated for reference.  

\placefigure{fig-zeroe}  
  
\placefigure{fig-separ}  

\subsection{General Considerations for Resonance Retention}\label{sec-gencon} 
 
The random component of the planet's migration is a form of Brownian motion. 
The planet encounters a large number of small planetesimals, each of which 
causes the planet's semi-major axis $a_{\rm p}$ to randomly step a small 
distance. 
Each random change in $a_{\rm p}$ produces a 
corresponding change in the semi-major axis of exact resonance,\footnote{A 
particle in exact resonance has zero libration amplitude, by definition.} 
but no  
corresponding change in the actual semi-major axis of a 
resonant particle. 
The semi-major axis of the particle does not respond 
because $a_{\rm p}$ changes randomly every encounter 
time $\Delta t_{\rm e}$, which is much shorter than the resonant libration 
period.
Only changes in $a_{\rm p}$ that are coherent over timescales 
longer than the libration period produce an adiabatic response in the 
particle's semi-major axis. We have verified these assertions by numerical 
orbit integrations (not shown). 
 
The no-response condition implies that as $a_{\rm p}$ random walks, 
the difference between the semi-major axes of exact resonance and of 
a resonant particle random walks correspondingly. 
In other words, a resonant particle's libration amplitude 
random walks. 
The sign and magnitude of each step in the libration amplitude's random walk 
depend on the phase of libration when the step is taken. 
Since at any given time an ensemble of 
resonant particles are distributed over the full range of phases, a single 
random-walk history for the planet generates an ensemble of different 
random-walk histories for the particles. 
 
When the libration amplitude of a resonant particle random walks past its 
maximum allowed value, the particle escapes resonance. The maximum libration 
amplitude (full width) as measured in semi-major axis is 
\begin{equation}\label{eqn-aplib}  
\delta a_{\rm p,lib} = 2C_{\rm lib} a_{\rm p}\left(\frac{M_{\rm p} e_{\rm res}}{M_*}\right)^{1/2} \,\, ,  
\end{equation}  
where $e_{\rm res}$ is the eccentricity of the resonant object 
(not to be confused with the planetesimals generating the bulk of the noise), 
$C_{\rm lib}\approx 4\sqrt{f_{31}/3}$ is a constant 
(see Murray and Dermott 1999 for $f_{31}$), 
and we have restricted consideration to first-order ($p-q=1$) resonances. 
For the 3:2 exterior resonance with Neptune, 
$C_{\rm lib}\approx 3.64$.  
Note that in contrast to the usual definition 
of maximum libration width, $\delta a_{\rm p,lib}$ refers not to the 
particle's semi-major axis, but rather to the planet's. 
The meaning of $\delta a_{\rm p,lib}$ is as follows. 
Take a particle in exact resonance. By definition, such a particle
has zero libration amplitude. Then 
the planet's semi-major axis can change instantaneously by at most 
$\delta a_{\rm p,lib}/2$ and the particle will still remain in resonance 
(but with finite libration amplitude).

Equation (\ref{eqn-aplib}) derives from the pendulum model of resonance,
which is known to be inaccurate at large
$e_{\rm res}$ for some resonances.
Malhotra (1996) finds numerically that for $e_{\rm res} = 0.1$--0.4,
$\delta a_{\rm p,lib}$ for the 3:2 resonance is insensitive to $e_{\rm res}$,
whereas the pendulum model predicts that $\delta a_{\rm p,lib}$ doubles over
this range. We nevertheless employ Equation (\ref{eqn-aplib})
to estimate the maximum
libration width, since it is simple, analytic, and introduces errors less than of
order unity in our numerical evaluations below. The qualitative
physics described in this
paper does not depend on the accuracy to which we estimate
$\delta a_{\rm p,lib}$.

Consider a planet which migrates outward on average. 
When the random component of the planet's migration is added to the average 
component, a planet can migrate either outward or inward at any moment. 
Call $S_{\rm rnd}  = \int_0^t \dot{a}_{\rm p,rnd} \, dt$
the running sum of the random changes in $a_{\rm p}$. 
The probability $P_{\rm keep}$ 
that a given particle is retained in resonance over some duration 
of migration equals the probability that $|S_{\rm rnd}|$ 
remains less than the maximum libration half-width $\delta a_{\rm p,lib}/2$ 
during that time. 
A particle that escapes resonance by being dropped behind the resonance
($S_{\rm rnd} = +\delta a_{\rm p,lib}/2$) is, practically speaking, permanently
lost. The planet cannot recapture the particle by smoothly migrating inward
(see \S\ref{sec-outin}). The random component of the planet's migration
can cause the planetesimal to be recaptured, but a recaptured particle
lies on a trajectory near the separatrix and quickly re-escapes in practice.
Once the average (outward) component of the planet's migration
carries the resonance well past the particle, the particle
cannot be recaptured even if $S_{\rm rnd}$ random walks back to zero;
in other words, the particle has been permanently left behind.
A particle that escapes by being dropped in front of the resonance
($S_{\rm rnd} = -\delta a_{\rm p,lib}/2$) is also lost more often than not. 
Such a particle can be recaptured when the planet resumes
migrating outward. Nevertheless, upon its recapture onto a trajectory near
the separatrix, the particle can librate back to smaller semi-major axes
and be expelled behind the resonance permanently.
 
\subsection{Order-of-Magnitude Planetesimal Sizes}\label{sec-oomsize} 
Armed with the considerations of \S\ref{sec-gencon}, we are now ready to 
derive analytic, order-of-magnitude expressions for the maximum planetesimal 
sizes compatible with resonance retention, for the simple case when the disk is composed
of objects of a single size.  The assumption of a single size is relaxed in \S\ref{sec-sizedist}.

Say the planet takes time $T$ to migrate at speed $\dot a_{\rm p,avg}$ from its initial to its final semi-major axis.  
Over this time, $a_{\rm p}$ random walks an expected distance of
\begin{eqnarray}\label{eqn-sigmaT}  
\sigma_{a_{\rm p},T} &\sim& \left<\dot a_{\rm p,rnd}^2\right>^{1/2}T \nonumber \\
&\sim&  
  \left\{\begin{array}{l}
     \displaystyle{\mathcal{C}\mathcal{R}^{-4} \left( \frac{\mathcal{M} m}{M_{\rm p}} \right)^{1/2} \frac{a_{\rm p}}{a_{\rm d}} \, e_{\rm H} \, v_{\rm H} \left(\frac{T}{\Omega_{\rm p}}\right)^{1/2}}, \\\qquad\qquad\qquad\qquad\rule{0in}{0.15in}\mbox{if $e<e_{\rm H}/\mathcal{R}^2$;} \\  \\
      \rule{0ex}{5ex}\displaystyle{\mathcal{C}\mathcal{R}^{-2} \left( \frac{\mathcal{M} m}{M_{\rm p}} \right)^{1/2}\frac{a_{\rm p}}{a_{\rm d}} \, e \, v_{\rm H} \left(\frac{T}{\Omega_{\rm p}}\right)^{1/2}},\\\qquad\qquad\qquad\qquad \rule{0in}{0.15in}\mbox{if $e_{\rm H}/\mathcal{R}^2 < e < \mathcal{R}e_{\rm H}$;} \\  \\
      \rule{0ex}{5ex}\displaystyle{\mathcal{C}\left( \frac{\mathcal{M} m}{M_{\rm p}} \right)^{1/2}\frac{a_{\rm p}}{a_{\rm d}}  \, \frac{e_{\rm H}^2}{e} \, v_{\rm H} \left(\frac{T}{\Omega_{\rm p}}\right)^{1/2}} , \\\qquad\qquad\qquad\qquad\rule{0in}{0.15in}\mbox{if $e > \mathcal{R}e_{\rm H}$,}  
  \end{array}\right.  
\end{eqnarray}  
where we have set $\Delta t=T$ in evaluating $\left<\dot a_{\rm p,rnd}^2\right>^{1/2}$.

If $\sigma_{a_{\rm p},T} < \delta a_{\rm p,lib}/2$, 
the planet can keep a large fraction of planetesimals in resonance.  
That is, most particles are retained in resonance when 
the disk mass comprises planetesimals of mass  
\begin{equation}\label{eqn-mlim}  
m \lesssim m_{\rm crit} \sim    
  \left\{\begin{array}{l}   
     \displaystyle{\frac{\mathcal{R}^8 C_{\rm lib}^2}{\mathcal{C}^2\mathcal{M}} \left(\frac{a_{\rm d}}{a_{\rm p}}\right)^2  \frac{1}{\Omega_{\rm p}T} \, \frac{e_{\rm res}}{e_{\rm H}} \, M_{\rm p}} , \\\qquad\qquad\qquad\qquad\rule{0in}{0.15in}\mbox{if $e<e_{\rm H}/\mathcal{R}^2$;} \\  \\
      \rule{0ex}{5ex}\displaystyle{\frac{\mathcal{R}^4 C_{\rm lib}^2}{\mathcal{C}^2\mathcal{M}} \, \left(\frac{a_{\rm d}}{a_{\rm p}}\right)^2  \frac{1}{\Omega_{\rm p}T} \frac{e_{\rm res}e_{\rm H}}{e^2} \, M_{\rm p}}, \\\qquad\qquad\qquad\qquad\rule{0in}{0.15in}\mbox{if $e_{\rm H}/\mathcal{R}^2 < e < \mathcal{R}e_{\rm H}$;} \\   \\
     \rule{0ex}{5ex}\displaystyle{\frac{C_{\rm lib}^2}{\mathcal{C}^2\mathcal{M}} \left(\frac{a_{\rm d}}{a_{\rm p}}\right)^2  \frac{1}{\Omega_{\rm p}T} \, \frac{e_{\rm res}e^2}{e_{\rm H}^3} \, M_{\rm p}}, \\\qquad\qquad\qquad\qquad\rule{0in}{0.15in}\mbox{if $e > \mathcal{R}e_{\rm H}$.}  
  \end{array}\right.   
\end{equation}   
Equation (\ref{eqn-mlim}) can be equivalently interpreted as an upper limit on  
$T$ for planetesimals of given mass $m$.  For a fixed degree of noise, resonant objects are more difficult to retain if the average migration is slow.  
  
We evaluate (\ref{eqn-mlim}) to estimate the maximum planetesimal radius,  
$s = (3m/4\pi\rho)^{1/3}$, compatible with resonant capture of KBOs by Neptune.   
For an internal density $\rho= 2\gm/{\rm cm}^3$, $M_{\rm p}=M_{\rm N}=17M_\earth$, $e_{\rm H} = 0.03$, $a_{\rm p} = a_{\rm d} = 26.6$ AU, $e_{\rm res} = 0.25$, $\mathcal{M} = 2$ (so that $\Sigma_m = 0.2 \gm \cm^{-2}$), and $T = 3\times 10^7$ yr, resonant capture and retention require  
\begin{equation}\label{eqn-smax} 
\frac{s}{{\rm km}} \, \lesssim \, \frac{s_{\rm crit}}{{\rm km}} \sim   
  \left\{\begin{array}{ll}  
    700\mathcal{R}^{8/3}\mathcal{C}^{-2/3}, &\mbox{if  $e<e_{\rm H}/\mathcal{R}^2$;} \\  
    \rule{0ex}{3ex} 70e^{-2/3}\mathcal{R}^{4/3}\mathcal{C}^{-2/3},  &\mbox{if $e_{\rm H}/\mathcal{R}^2 < e < \mathcal{R}e_{\rm H}$;} \\  
    \rule{0ex}{3ex} 7000e^{2/3}\mathcal{C}^{-2/3}, &\mbox{if $e > \mathcal{R}e_{\rm H}$.}  
  \end{array}\right.  
\end{equation}  
For example, if $e = 0.1$ and $\mathcal{R} = 1$, then line 3 of (\ref{eqn-smax}) obtains and $s_{\rm crit} = 1500 \mathcal{C}^{-2/3} \km$.  Maximum stochasticity results when $\mathcal{R} = 1$ and $e \leq e_{\rm H}$ (see also Equation [\ref{eqn-maxstoch}]); either of lines 1 or 2 then yield 
$s_{\rm crit} \sim 700\mathcal{C}^{-2/3} \km$. These size estimates decrease
by about 20\% when corrected to reflect the fact that the width of the 3:2 resonance
is somewhat smaller than the pendulum model implies (see the discussion
following Equation [\ref{eqn-aplib}]).
  
\subsection{Analytical Formula for the Retention Fraction}\label{sec-analyt}  
 
As defined in \S\ref{sec-gencon}, $P_{\rm keep}$ is the resonance 
retention fraction, or the probability 
that a typical resonant particle is retained in resonance over   
some duration of migration. We calculate $P_{\rm keep}$ 
by modelling the random 
component of the planet's migration as a diffusive continuum process. 
In the limit that the planet encounters a large number $\overline{N} \gg 1$ 
of planetesimals, the Poisson distribution 
(Equation [\ref{eqn-poiss}]) is well-approximated by a Gaussian 
distribution with mean $\overline{N}$ and variance $\overline{N}$.   
Thus, over a time interval $\Delta t \gg \dot{\overline{N}}^{-1}$,
the random displacement of the planet, 
$\Delta S_{\rm rnd} = \Delta a_{\rm p}(N-\overline{N})$, 
has the probability density distribution  
\begin{equation}\label{eqn-disppdf}  
f(\Delta S_{\rm rnd},\Delta t) = \frac{1}{\sqrt{2\pi D\Delta t}} \exp (-(\Delta S_{\rm rnd})^2/(2D\Delta t)) \,\, ,  
\end{equation}  
where $D = (\Delta a_{\rm p})^2\dot{\overline{N}}$ is the diffusion 
coefficient and we recall that $\Delta a_{\rm p}$ is the change in $a_{\rm p}$ 
due to an encounter with a single planetesimal. 
The evolution of $a_{\rm p,rnd}$ with $t$ 
is continuous and the distribution $f$ is independent over any two 
non-overlapping intervals $\Delta t$ (the random walk has no memory). 
In other words, $\Delta S_{\rm rnd}$ evolves  
as a Wiener process, or equivalently according to the rules of Brownian motion (e.g., Grimmett \& Stirzaker 2001a). 
From Equation (\ref{eqn-disppdf}), it follows that over 
time $T$, the probability that 
$|\Delta S_{\rm rnd}|$ does not exceed $\delta a_{\rm p,lib}/2$ equals 
\begin{equation}\label{eqn-abs}  
P_{\rm keep} = \sum_{n=1}^{\infty} \frac{4}{n\pi}\sin{^3\left(\frac{n\pi}{2}\right)}e^{-\lambda_n T} \,\, ,  
\end{equation}  
where $\lambda_n = (n\pi)^2D/(2\delta a_{\rm p,lib}^2)$ (see Appendix \ref{ap-abs} for a derivation).

\begin{figure*}  
\epsscale{1.0}  
\plotone{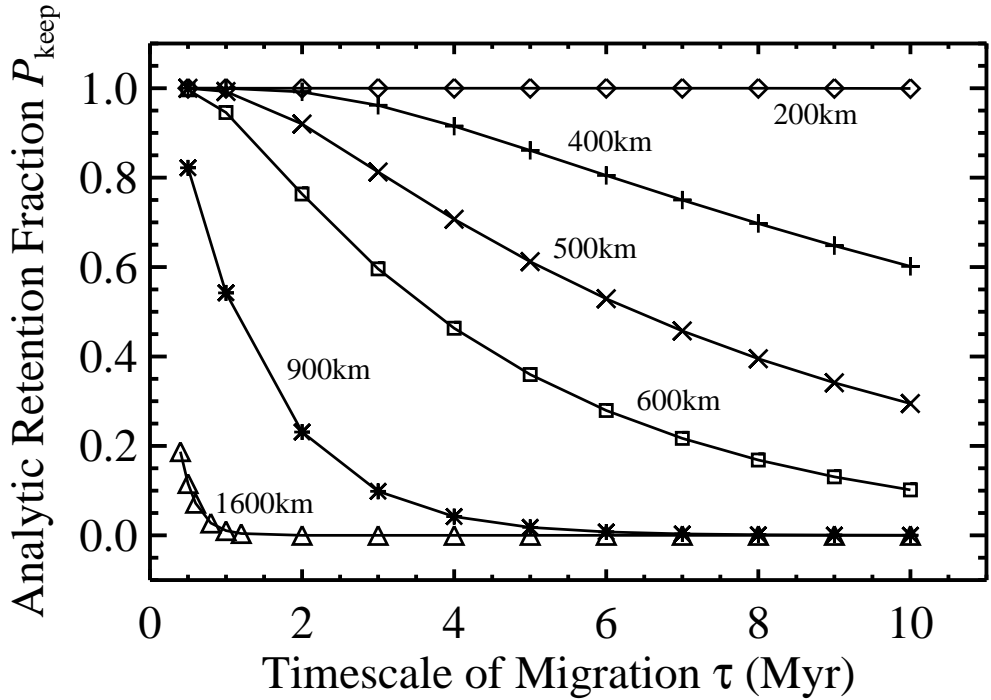}  
\caption{Fraction of particles retained in external 3:2 resonance by  
a stochastically migrating planet as a 
function of migration timescale, 
calculated according to Equation (\ref{eqn-abs}). The entire disk mass is assumed to be in planetesimals of a single size $s$, and a range of choices for $s$ are shown.  The diffusivity $D$ is 
evaluated at its maximum value, appropriate for the case $e \leq e_{\rm H}$ 
and $\mathcal{R} = 1$. We set $\Delta a_{\rm p} = \Delta a_{\rm p,nc1}$ 
(Equation [\ref{eqn-sub}]), 
$\dot{\overline{N}} = 2 \Sigma_m \Omega R_{\rm H}^2 / m$, 
(Equation [\ref{eqn-subencounter}]), 
$\rho= 2\gm/{\rm cm}^3$, 
$M_{\rm p}=M_{\rm N}=17M_\earth$, $e_{\rm H} = 0.03$, 
$a_{\rm p} = a_{\rm d} = 26.6$ AU, 
$R_{\rm H} = e_{\rm H}a_{\rm p}$, $\mathcal{M} = 2$, 
$e_{\rm res} = 0.25$, and $T = 2.6 \tau$. 
Planetesimals having sizes smaller than $\sim$200 km produce so little 
noise in the planet's migration 
that no object is lost from the 3:2 resonance. Compare this Figure with 
its numerical counterpart, Figure \ref{fig-rates}. In calculating $P_{\rm keep}$, we assume $\mathcal{C} = 1$; probably $\mathcal{C}$ is of order several, in which case the sizes indicated in the Figure should be revised downward by a factor of a few ($\mathcal{C}^{2/3}$; see Equation [\ref{eqn-smax}]).
}  
\label{fig-analytrates}  
\end{figure*}

Suppose migration occurs in a disk of planetesimals having a single size $s$.  Figure \ref{fig-analytrates} displays $P_{\rm keep}$ as a function 
of $s$ and of exponential migration timescale $\tau$ defined 
according to 
\begin{equation}\label{eqn-aavg} 
a_{\rm p,avg}(t) = a_{\rm p,f} - (a_{\rm p,f} -  a_{\rm p,i})e^{-t/\tau} \,, 
\end{equation}  
where $a_{\rm p,i}$ and $a_{\rm p,f}$ are the planet's initial and final 
average semi-major axes, respectively. In Equation (\ref{eqn-abs}), 
we take $T = 2.6\tau$, and evaluate remaining quantities 
for the case of maximum stochasticity: $e \leq e_{\rm H}$ 
and $\mathcal{R} = 1$. Then $\Delta a_{\rm p} = \Delta a_{\rm p,nc1}$ 
(Equation [\ref{eqn-sub}]) and 
$\dot{\overline{N}} = 2 \Sigma_m \Omega R_{\rm H}^2 / m$ 
(Equation [\ref{eqn-subencounter}], with a factor of 2 inserted 
to account for disk material both inside and outside the planet's 
orbit). As in \S\ref{sec-gencon}, we take $\rho= 2\gm/{\rm cm}^3$, 
$M_{\rm p}=M_{\rm N}=17M_\earth$, $e_{\rm H} = 0.03$, 
$a_{\rm p} = a_{\rm d} = 26.6$ AU, 
$R_{\rm H} = e_{\rm H}a_{\rm p}$, and $\mathcal{M} = 2$. 
To evaluate $\delta a_{\rm p,lib}$, we take $e_{\rm res} = 0.25$ 
for a particle in 3:2 resonance. Figure 
\ref{fig-analytrates} describes how for a given size $s$, the retention 
fraction decreases with increasing $\tau$; the longer the duration 
of migration, the more chance a particle has of being jostled out of 
resonance. For $\tau = 10$ Myr, planetesimals must have sizes 
$s \lesssim 500 \km$ 
for the retention fraction to remain greater 
than 1/2.  
These results confirm and refine our order-of-magnitude 
estimates made in \S\ref{sec-oomsize}.   
Similar results were obtained for the 2:1 resonance. 
 
\placefigure{fig-analytrates}  
The continuum limit is valid as long as 
the expectation value of the time required for a resonant particle to escape, 
$\left< t_{\rm escape} \right> \sim \dot{\overline{N}}^{-1} 
\left[ (\delta a_{\rm p,lib}/2)/\Delta a_{\rm p} \right]^2$, 
greatly exceeds the time for the planet to encounter one planetesimal, 
$\dot{\overline{N}}^{-1}$. 
This criterion is satisfied for the full range of parameters adopted 
in Figure \ref{fig-analytrates}.

\subsection{Numerical Results for the Retention Fraction} \label{sec-num}  
  
To explore how a stochastically  
migrating planet captures and retains 
test particles into its exterior resonances, and to test the 
analytic considerations of \S\S\ref{sec-gencon}--\ref{sec-analyt}, 
we perform a series of numerical integrations.  
We focus as before on the 3:2 (Plutino) resonance with Neptune. 
  
Following  
Murray-Clay \& Chiang (2005, hereafter MC05), we employ a series expansion  
for the time-dependent Hamiltonian,  
\begin{eqnarray}\label{eqn-hamiltonian}
H &=& -\frac{(G M_{\odot})^2}{2(3\Gamma+\mathcal{N})^2} - \left[\frac{GM_{\odot}}{a_{\rm p}(t)^3}\right]^{1/2} \left( 2\Gamma+\mathcal{N} \right) \nonumber \\
&&- \frac{GM_{\rm p}}{a_{\rm p}(t)} \left[ \alpha (f_{1} + f_{2}e^2 + f_{31} e\cos \phi) \right] \,,  
\end{eqnarray}  
\noindent where $\alpha = a_{\rm p}/a \approx 0.76$, the $f_{i}$'s are given  
in Murray \& Dermott (1999), and $\mathcal{N}$ (Equation [\ref{eqn-Nadiabat}]) 
is a constant of the motion  
determined by initial conditions. The resonance angle,  
\begin{equation}\label{eqn-resangle}
\phi = 3 \lambda_{\rm res} - 2 \lambda_{\rm p} - \pomega_{\rm res} \,,  
\end{equation}  
is defined by the mean longitude $\lambda_{\rm res}$ and longitude of  
periastron $\pomega_{\rm res}$ of the resonant particle,  
and the mean longitude $\lambda_{\rm p}$ of the planet.  
The resonance angle librates about $\pi$ for particles  
in resonance. The momentum conjugate to $\phi$ is $\Gamma$.  
We integrate the equations of motion,  
\begin{equation}  
\dot{\phi} = \frac{\partial H}{\partial \Gamma} \,, \,\,\,\, \dot{\Gamma} = -\frac{\partial H}{\partial \phi} \,,  
\end{equation}  
using the Bulirsch-Stoer algorithm (Press et al.~1992) for fixed $\alpha$  
and $f_{i}$'s.  

The Hamiltonian in Equation (\ref{eqn-hamiltonian})
faithfully reproduces the main features of the
resonance potential; see Beaug\'{e} (1994, his Figures 12a and 12c) for
a direct comparison between such a truncated Hamiltonian and the exact
Hamiltonian, averaged over the synodic period  (see also that paper and Murray-Clay \& Chiang 2005 for a
discussion of the pitfalls of keeping one too many a term in the expansion).
Of course, even the exact Hamiltonian, because it is time-averaged and neglects chaotic zones, is
inaccurate with regards to details such as the libration width,
but these inaccuracies are slight; see the discussions following
Equations (\ref{eqn-aplib}) and (\ref{eqn-smax}).
  
To compute $a_{\rm p}(t)$, we specify separately 
the average and random components of the migration velocity, 
$\dot a_{\rm p,avg}$ and $\dot a_{\rm p,rnd}$. 
For $\dot a_{\rm p,avg}$, we adopt the prescription (equivalent to Equation [\ref{eqn-aavg}]) 
\begin{equation}\label{eqn-adotavg}  
\dot a_{\rm p,avg} = \frac{1}{\tau}(a_{\rm p,f} -  a_{\rm p,i})e^{-t/\tau} \,\, ,  
\end{equation}  
where $a_{\rm p,i}$ and $a_{\rm p,f}$ are the planet's initial and final 
average semi-major axes, respectively, and $\tau$ is a time constant. 
To compute $\dot a_{\rm p,rnd}$, we divide the integration  
into time intervals of length $1/\Omega_{\rm p}$.   
The only requirement for the time interval is that
it be less than the libration period $t_{\rm lib} \sim 400/\Omega_{\rm p}$ (see
\S\ref{sec-gencon}).
Over each interval, we randomly generate 
\begin{equation}\label{eqn-adotrnd}  
\dot a_{\rm p,rnd} = \Omega_{\rm p} \Delta a_{\rm p} (N_{\Omega}-\dot{\overline N}\Omega_{\rm p}^{-1}) \,\, . 
\end{equation}  
We focus on the case of maximum stochasticity, so that 
$\Delta a_{\rm p} = \Delta a_{\rm p,nc1}$ (Equation [\ref{eqn-sub}]) 
and $\dot{\overline N} = 2\Sigma_m R_{\rm H}^2 \Omega_{\rm p}/m$ 
(Equation [\ref{eqn-subencounter}] with $\mathcal{R} = 1$ and 
an extra factor of 2 inserted to account for disk material 
on both sides of the planet's orbit).  We assume that the entirety of the disk mass is in planetesimals of a single mass $m$.
Each $N_{\Omega}$ is a random deviate drawn from a Poisson distribution 
having mean $\dot{\overline N} \Omega_{\rm p}^{-1}$.

\begin{figure}  
\epsscale{1.2}  
\plotone{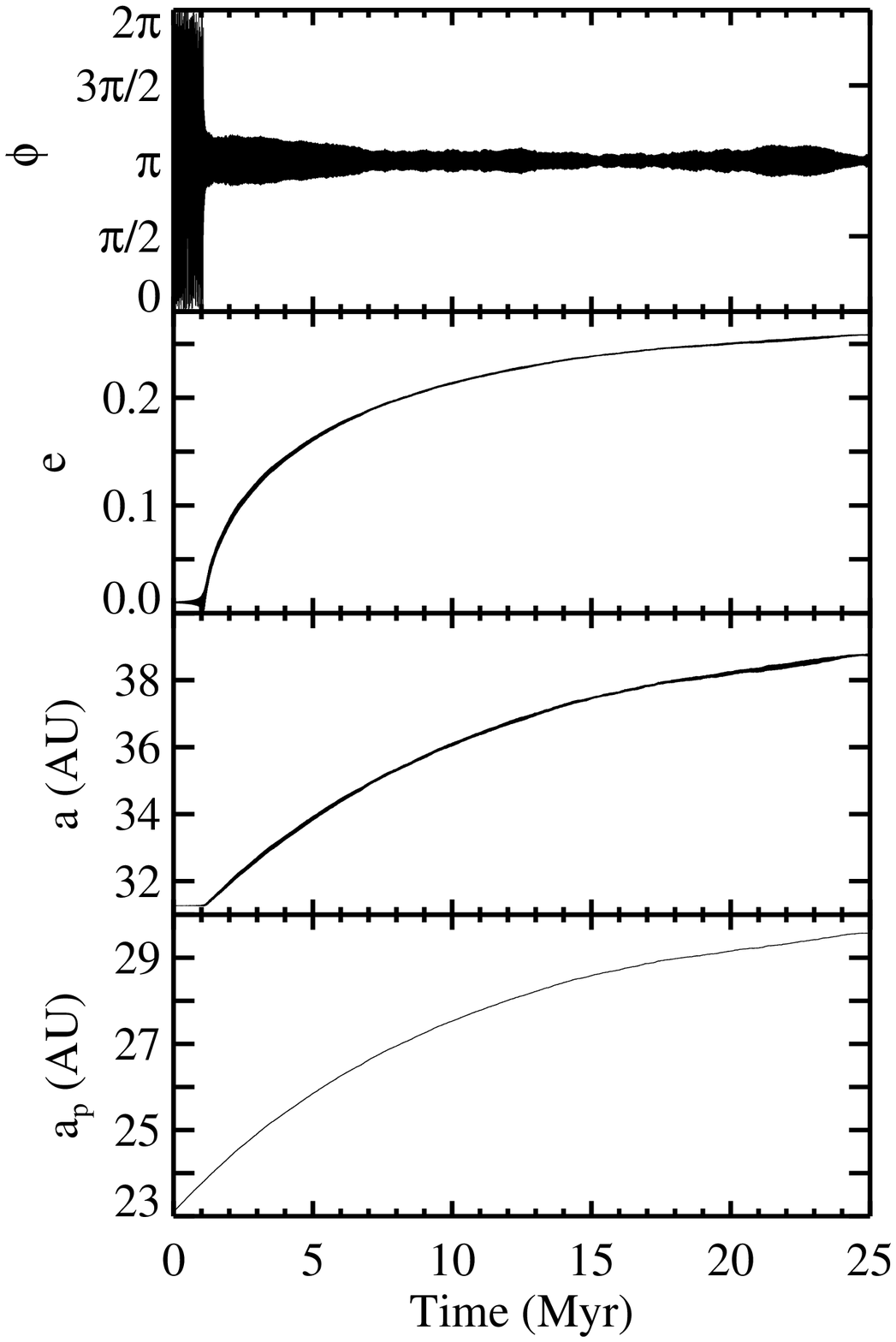}  
\caption{Evolution of a particle caught into 3:2 resonance 
with a stochastically migrating planet. Stochasticity is driven by 
a disk of surface density $\Sigma_m = 0.2 \gm \cm^{-2}$, all in planetesimals having sizes $s = 150 \km$ 
and sub-Hill random velocities. The random walk in the planet's semi-major 
axis causes the libration amplitude of the resonant particle to undergo 
a corresponding random walk. 
The noise in this example is too mild to prevent 
the planet from both capturing and retaining the particle in resonance. 
}  
\label{fig-apsubsm1}  
\end{figure}

Figure \ref{fig-apsubsm1} displays the sample evolution of a test particle  
driven into 3:2 resonance by a stochastically migrating planet.  
For this integration,  $M_{\rm p} = M_{\rm N}$, $a_{\rm p,i} = 23.1\AU$, 
$a_{\rm p,f} = 30.1\AU$, $\tau = 10^7 \yr$,  
$\mathcal{R} = 1$, 
$\mathcal{M} = 2$ 
(so that $\Sigma_m = 0.2 \gm \cm^{-2}$), 
and $s = 150\km$ 
($m = 3\times10^{22}\gm$). 
This choice for $s$ is sufficiently  
small that the particle is successfully captured and retained  
by the planet. 
 
\begin{figure}  
\epsscale{1.2}  
\plotone{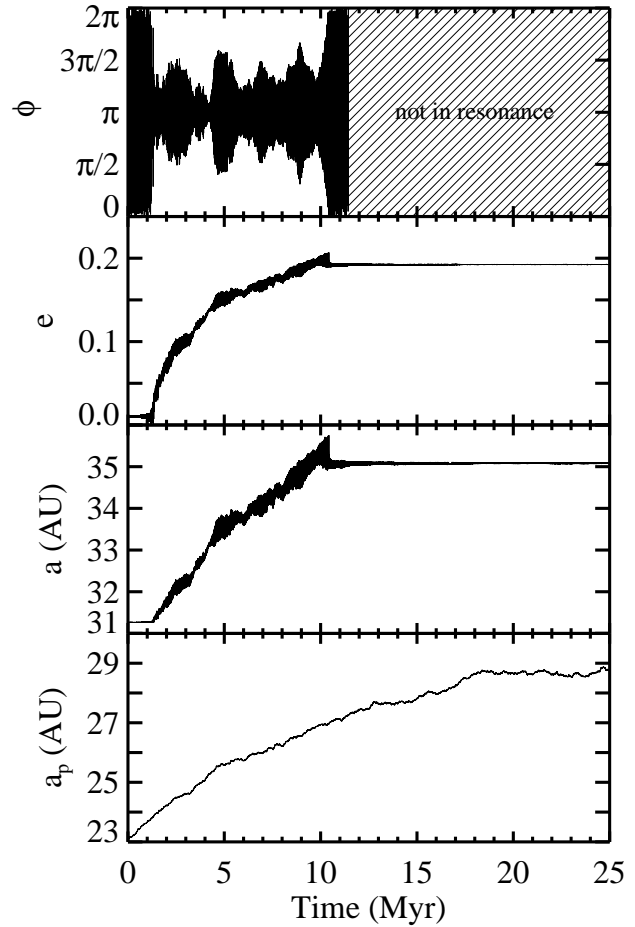}  
\caption{Evolution of a particle caught into, but eventually lost from, 
3:2 resonance with a stochastically migrating planet. Stochasticity is driven 
by a disk of surface density $\Sigma_m = 0.2 \gm \cm^{-2}$, all in planetesimals having sizes $s = 700 \km$ and sub-Hill random 
velocities. The particle is expelled from resonance having had its 
eccentricity raised to $0.2$ during its time in resonant lock. 
} 
\label{fig-apsublg1}  
\end{figure}

Contrast Figure \ref{fig-apsubsm1} with Figure \ref{fig-apsublg1}, in which  
all model parameters are the same except for a larger $s = 700\km$.  
In this case the planet eventually loses the test particle because  
the migration is too noisy.  
  
\begin{figure*}  
\epsscale{1.0}  
\plotone{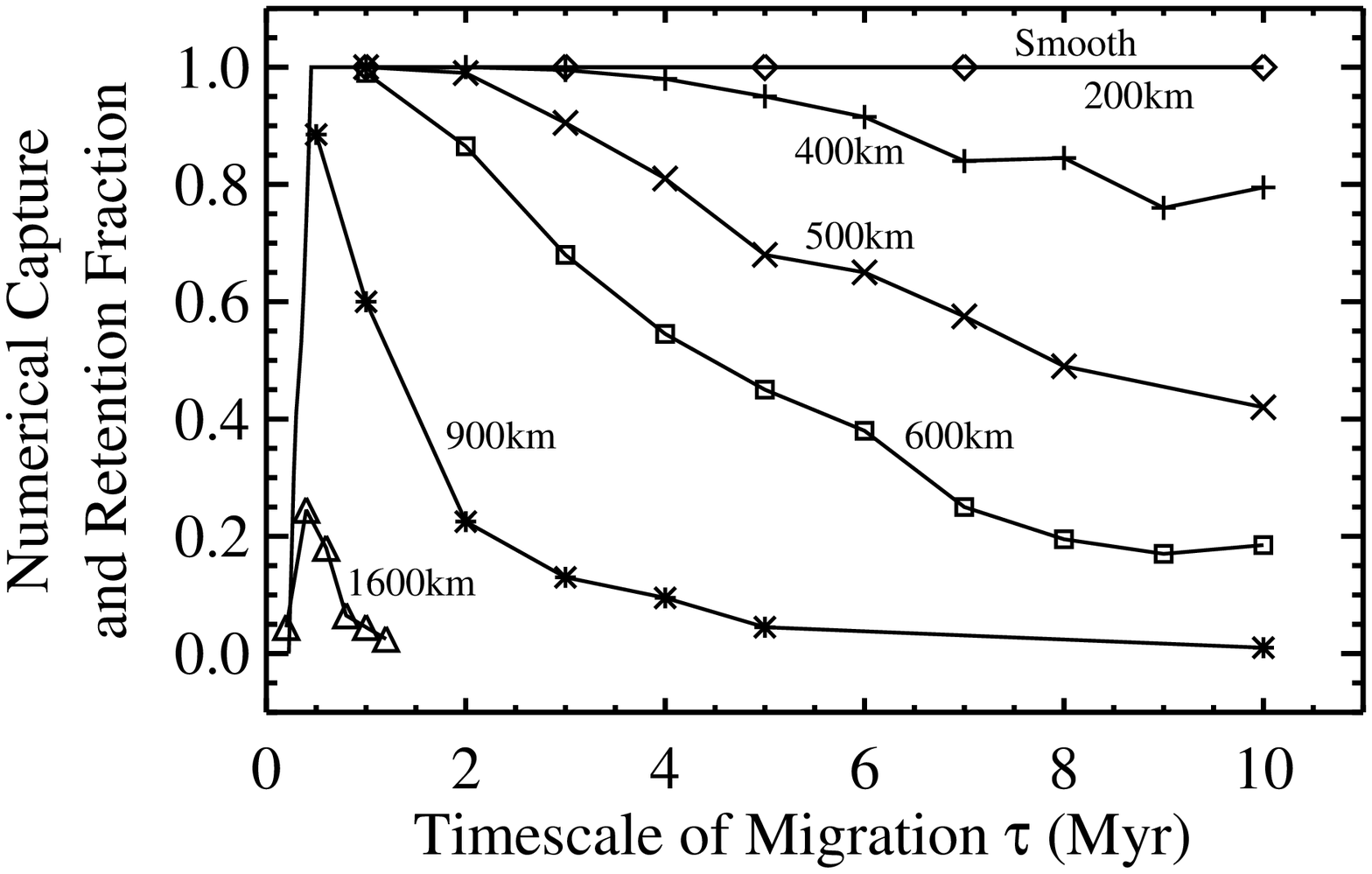}  
\caption{Fraction of particles caught into, and retained within, 
external 3:2 resonance by a stochastically migrating planet. 
For every $\tau$ and $s$, we numerically integrate 
the trajectories of 200 test particles with initial eccentricities 
of $\sim$$0.01$ and semi-major axes that lie 1 AU outside of nominal 
resonance. These particles respond to the time-averaged potential of 
a Neptune-mass planet which migrates outward from $23.1\AU$ to $30.1\AU$ 
within a disk of fixed surface density $\Sigma_m = 0.2 \gm \cm^{-2}$ in planetesimals of a single size $s$. The planetesimals have sub-Hill random velocities 
and semi-major axes that lie within $\mathcal{R} = 1$ Hill radius of 
the planet's; these choices maximize the amount of stochasticity 
in the planet's migration. Compare this Figure with its analytic counterpart, 
Figure \ref{fig-analytrates}; the agreement is excellent. 
The solid curve labelled ``Smooth'' corresponds to the case 
when all noise is eliminated from the planet's migration. Planetesimals 
having sizes smaller than $\sim$$200 \km$ yield an essentially smooth 
migration. For $\tau \lesssim 10^5 \yr$, capture is not possible even 
if migration were smooth, since the migration is too fast to be adiabatic.  These results are calculated for $\mathcal{C} = 1$; probably $\mathcal{C}$ is of order several and so the sizes indicated in the Figure should be revised downward by a factor of a few ($\mathcal{C}^{2/3}$; see Equation [\ref{eqn-smax}]). 
} 
\label{fig-rates}  
\end{figure*}

Figure \ref{fig-rates} displays the fraction of particles caught and kept  
in resonance as a function of $\tau$ and $s$. 
For each data point in Figure \ref{fig-rates}, we follow the evolution 
of 200 particles initialized with   
eccentricities of approximately 0.01 and semi-major axes  
that lie outside the initial position of resonance by  
about 1 AU.\footnote{The particles  
do not all have the same initial eccentricities and semi-major axes.  This is 
because they occupy the same Hamiltonian level curve;  
see section 3.5 of MC05.} 
Figure \ref{fig-rates} is the numerical counterpart of 
Figure \ref{fig-analytrates}; the agreement between the two 
is excellent and validates our analytic considerations. 
If $\tau = 10^7 \yr$ (consistent with findings by MC05),  
then the capture fraction rises above 0.5 for $s \lesssim 500 \km$.  
Since these results pertain to the case \{$\mathcal{R} = 1$, 
$e \leq e_{\rm H}$\} which yields  
the largest amount of noise for given $\Sigma_m$ and $s$, we conclude 
that $s \sim 500 \km$ is the lowest, and thus the most conservative, estimate we can make  
for the maximum planetesimal size compatible with resonant capture of KBOs by a migrating Neptune, assuming that the entire disk is composed of planetesimals of a single size (this assumption is relaxed in the next section).
In other words, if Neptune's migration  
were driven by planetesimals all having $s \ll 500 \km$,  
stochasticity would not have impeded the trapping of Resonant KBOs.  
Of course, our numerical estimate of $500 \km$ is uncertain insofar 
as we have not kept track of order-unity constants in our derivations. 
We suspect a more careful analysis will revise our size estimate downwards 
by a factor of a few (see the discussion of $\mathcal{C}$ in 
\S\ref{sec-neatsum}). 
 
\placefigure{fig-apsubsm1}  
  
\placefigure{fig-apsublg1}  
  
\placefigure{fig-rates}  
  
\subsection{Planetesimal Size Distributions}
\label{sec-sizedist}

Actual disks comprise planetesimals with a range of sizes. From Equation
(\ref{eqn-rosetta}), the stochasticity in the planet's migration is
dominated by those planetesimals having maximal $\Sigma_mm$. What was
the distribution of sizes during the era of Neptune's migration?
A possible answer is provided by the coagulation simulations of
Kenyon \& Luu (1999, hereafter KL99). The left-hand panel of their
Figure 8 portrays the evolution of the size distribution,
starting with a disk of seed bodies having sizes up to 100 m
and a total surface density of $\Sigma = 0.2\gm\cm^{-2}$.
After $t = 11$ Myr, the size bin for which $\Sigma_mm$ is maximal
is centered at $s \sim 4 \km$; for this bin at that time,
$\Sigma_m = 10^{-3} \gm \cm^{-2}$ (evaluated within a logarithmic size
interval 0.3 dex wide). After $t = 37$ Myr, the planetesimals generating
the most stochasticity have $s\sim 750\km$ and
$\Sigma_m = 2\times 10^{-3} \gm \cm^{-2}$.
Note that at $t = 37$ Myr, the stochasticity is dominated by the largest planetesimals formed, but they do not contain the bulk of the total disk
mass; the lion's share of the mass is instead sequestered into
km-sized objects.

\begin{figure*}  
\epsscale{1.0} 
\plotone{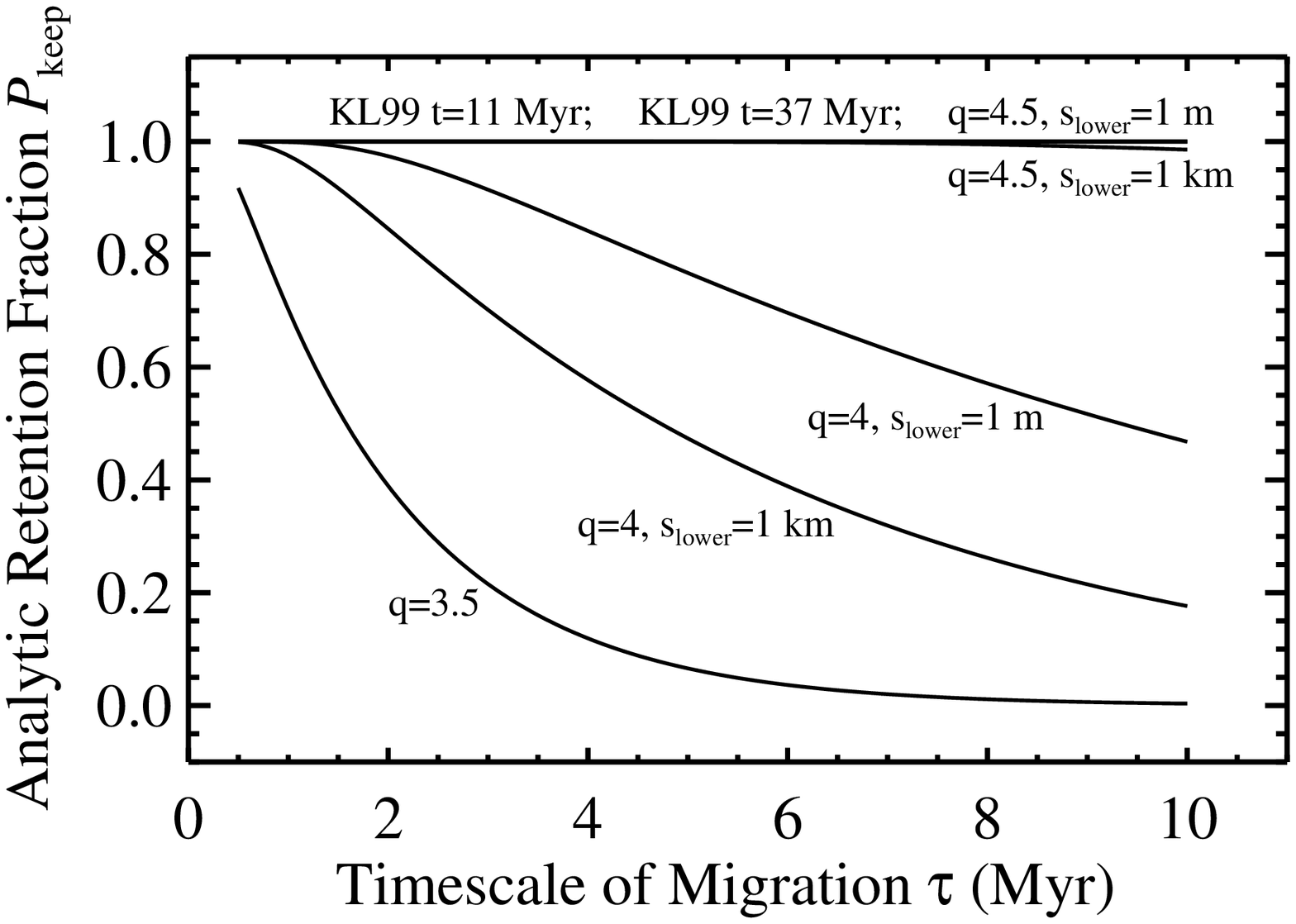}  
\caption{Fraction of particles retained in external 3:2 resonance by  
a stochastically migrating planet for various planetesimal size
distributions. The retention efficiency is calculated analytically
using Equation (\ref{eqn-abs}), with parameters the same as those
for Figure \ref{fig-analytrates} except for $\Sigma_m \times m$;
that parameter is evaluated at its maximum value
within a logarithmic size bin spanning a factor of 2
for a given size distribution.
The size distributions considered include two from Kenyon \& Luu (1999;
their Figure 8), evaluated at times $t = 11$ Myr and 37 Myr;
and five different power-law distributions, each characterized
by a total integrated surface density $\Sigma = 0.2 \gm \cm^{-2}$,
an upper size limit
$s_{\rm upper} = 1200 \km$,
a differential size index $q$ (such that $d\eta/ds\propto s^{-q}$), and a lower size limit $s_{\rm lower}$ as indicated
(the curve for
$q = 3.5$ is insensitive to $s_{\rm lower}$ since the bulk of the
mass is concentrated towards $s_{\rm upper}$). The three curves for the
size distributions of KL99 and for $\{q=4.5$, $s_{\rm lower} = 1\m\}$
overlap at $P_{\rm keep} = 1$.
}  
\label{fig-sizedist}
\end{figure*}

In Figure \ref{fig-sizedist}, we plot the resonance retention fraction
$P_{\rm keep}$ (Equation [\ref{eqn-abs}]) for the KL99 size distribution
at $t = 11$ and 37 Myr, using the values of $s(m)$ and $\Sigma_m$ cited
above. The remaining parameters that enter into $P_{\rm keep}$ are chosen
to be the same as those employed for Figure \ref{fig-analytrates}; i.e.,
we adopt the case of maximum stochasticity. Evidently, $P_{\rm keep} = 1$
for the KL99 size distributions; stochasticity is negligible.

For comparison, we also plot in Figure \ref{fig-sizedist} the retention
fraction for pure power-law size distributions: $d\eta/ds \propto s^{-q}$,
where $d\eta$ is the differential number of planetesimals having sizes between
$s$ and $s+ds$. Since $\Sigma_mm \propto s^{7-q}$, stochasticity
is dominated by the upper end of the size distribution for $q < 7$.
We fix the maximal radius to be that of Pluto
($s_{\rm upper} = 1200 \km$),
set the total surface density $\Sigma = 0.2 \gm \cm^{-2}$,
and calculate $P_{\rm keep}$ for three choices of $q = 3.5$, 4, and 4.5.
For $q \geq 4$, the lower limit of the size distribution significantly influences
the normalization of $d\eta/ds$; for $q = 4$ and 4.5, we experiment
with two choices for the minimum planetesimal radius,
$s_{\rm lower} = 1 \km$ and $1 \m$. We equate $\Sigma_m$ with the integrated
surface density between $s_{\rm upper}/2$ and $s_{\rm upper}$.
According to Figure \ref{fig-sizedist}, steep size distributions $q \geq 4$
are characterized by order-unity retention efficiencies.
In contrast, shallow size distributions $q < 4$ for which the bulk
of the mass is concentrated towards $s_{\rm upper}$ can introduce
significant stochasticity.

\placefigure{fig-sizedist}  

\section{CONCLUDING REMARKS}\label{sec-sum} 
  
We summarize our findings in \S\ref{sec-sum1} and discuss
quantitatively some remaining issues in \S\ref{sec-extend}.

\subsection{Summary}\label{sec-sum1}                          

Newly formed planets likely occupy remnant planetesimal disks.
Planets migrate as they exchange energy and angular momentum with
planetesimals.  Driven by discrete scattering events, migration is
stochastic.
 
In our solar system, Neptune may have migrated outward by several AU 
and thereby captured 
the many Kuiper belt objects 
(KBOs) found today in mean-motion resonance with the planet.     
While resonance capture is efficient when migration is smooth, 
a longstanding issue has been whether Neptune's actual migration was 
too noisy to permit capture.  Our work addresses---and dispels---this concern 
by supplying a first-principles theory for how a planet's semi-major axis 
fluctuates in response to intrinsic granularity in the gravitational 
potential. We apply our theory to identify the environmental conditions 
under which resonance capture remains viable. 
  
Stochasticity results from random variations in the 
numbers and orbital properties of planetesimals encountering the planet.   
The degree of stochasticity (as measured, say, by $\sigma_{a_{\rm p},T}$, 
the typical distance that the planet's semi-major axis random walks away from its average value) depends on how planetesimal semi-major axes $a$ 
and random velocities $u$ are distributed. We have parameterized $a$ by its 
difference from the planet's semi-major axis: 
$x \equiv a-a_{\rm p} \equiv \mathcal{R} R_{\rm H}$, where $R_{\rm H}$ 
is the Hill sphere radius and $\mathcal{R} \gtrsim 1$. 
In the case of high dispersion when $u > \mathcal{R} v_{\rm H}$ (where 
$v_{\rm H} \equiv \Omega_{\rm p}R_{\rm H}$ is the Hill velocity 
and $\Omega_{\rm p}$ is the planet's orbital angular velocity), 
planetesimal orbits cross that of the planet. 
Stochasticity increases with decreasing $u$ 
in the high-dispersion case because the cross-section for 
strong scatterings increases steeply with decreasing velocity 
dispersion (as $1/u^4$). 
In the intermediate-dispersion case when 
$v_{\rm H}/\mathcal{R}^2 < u < \mathcal{R}v_{\rm H}$, 
planetesimal and planet orbits do not cross, and stochasticity 
decreases with decreasing $u$. In the low-dispersion case when 
$u < v_{\rm H}/\mathcal{R}^2$, 
the amount of stochasticity is insensitive to $u$. 
 
The values of $u$ and $\mathcal{R}$ which actually characterize disks are 
unknown. The random velocity $u$, for example, is expected to be set by 
a balance between excitation by gravitational scatterings and damping 
by inelastic collisions between planetesimals and/or gas drag. Damping 
depends, in turn, 
on the size distribution of planetesimals. These considerations are 
often absent from current N-body simulations of planetary migration 
in planetesimal disks. Despite such uncertainty, 
we can still identify the circumstances under 
which stochasticity is maximal. Maximum stochasticity obtains when 
$\mathcal{R} \sim 1$ and $u \lesssim v_{\rm H}$, that is, when 
planetesimals have semi-major axes within a Hill radius of the planet's 
and when their velocity dispersion is no greater than the Hill velocity. 
 
A stochastically migrating planet cannot retain objects in a given 
resonance if the planet's semi-major axis random walks away from its average value by a distance greater than the 
maximum libration width of the resonance. This simple 
criterion is validated by numerical experiments and enables 
analytic calculation of the resonance retention efficiency as a function 
of disk parameters. A disk of given surface density generates more 
noise when composed of fewer, larger planetesimals. In the context of 
Neptune's migration, we estimate that if 
the bulk of the minimum-mass disk resided in bodies having sizes 
smaller than $\mathcal{O} (100)$ km and if the fraction of the disk mass in larger bodies was not too large ($\lesssim$ a few percent for planetesimals having sizes of $1000 \km$, for example), then the retention efficiency of 
Neptune's first-order resonances 
would have been of order unity ($\gtrsim 0.1$).
Such order-unity efficiencies seem
required by observations, which {\it prima facie} place 122/474 $\approx$ 26\%
of well-observed KBOs (excluding Centaurs) inside mean-motion resonances
(Chiang et al.~2006). Drawing conclusions based on a comparison between
this observed percentage and our
theoretical retention percentage $P_{\rm keep}$ is a task
fraught with caveats---a more fair comparison would require, e.g.,
disentangling the observational bias against discovering
Resonant vs.~non-Resonant objects;
account of the attrition of the Resonant population due to weak chaos over
the four-billion-year age of the solar system; and knowledge of the initial
eccentricity and semi-major axis distributions of objects prior to resonance
sweeping, as these distributions impact capture probabilities in different
ways for different resonances (Chiang et al.~2003; Hahn \& Malhotra 2005;
Chiang et al.~2006).
But each of these caveats alters the relevant percentages only by factors of
a few, and when combined, their effects tend to cancel. Therefore we feel
comfortable in our assessment that $P_{\rm keep}$ must have been of order
unity to explain the current Resonant population. In that case,
$\mathcal{O} (100\km)$ is a conservative estimate for the maximum allowed size of planetesimals comprising the bulk of the disk mass, derived for the case of maximum stochasticity.

How does an upper limit of $\mathcal{O}(100)\km$ compare with the actual
size distribution of the planetesimal disk? While today's Kuiper belt places
most of its mass in objects having sizes of $\sim$100 km, this total mass
is tiny---only $\sim$$0.1 M_{\earth}$ (Bernstein et al.~2004; see
Chiang et al.~2006 for a synopsis). The current belt is therefore 2--3
orders of magnitude too low in mass to have driven Neptune's migration.
The current size distribution is such that bodies having radii $\gtrsim 40\km$
are collisionless over the age of the solar system and might therefore
represent a direct remnant, unadulterated by erosive collisions, of the
planetesimal disk during the era of migration (Pan \& Sari 2005). If so, the
bulk of the primordial disk mass must have resided in bodies having sizes
$\lesssim 40 \km$. Theoretical calculations of the coagulation history of the
Kuiper belt are so far
consistent with this expectation. Kenyon \& Luu (1999) find, for their
primordial trans-Neptunian disk of $10 M_{\earth}$, that 99\% of the mass
failed to coagulate into bodies larger than $\mathcal{O}(1)\km$, because
the formation of several Pluto-sized objects (comprising $\sim$0.1\% of the
total mass) excited velocity dispersions so much that planetesimal collisions
became destructive rather than agglomerative.  The average-mass planetesimals in their simulation have sizes $\mathcal{O}(1) \km$, much smaller than even our most conservative estimate of the maximum
allowed size of $\mathcal{O}(100)\km$.

For a given size distribution of planetesimals, most stochasticity is produced by the size bin having maximal $\eta\, m^2$, which need not be the size bin containing the majority of the mass. Here, $\eta$ and $m$ are the number of planetesimals and the mass of an individual planetesimal in a logarithmic size bin. For power-law size distributions $d\eta/ds \propto s^{-q}$ such that $q < 7$, stochasticity is dominated by the largest planetesimals.  
For disks having as much mass as the minimum-mass disk of solids and whose largest members are Pluto-sized, size distributions with $q \ge 4$ enjoy order-unity efficiencies for resonance retention.
The size distributions of Kenyon \& Luu (1999) resemble $q=4$ power laws, but with a large overabundance of planetesimals having sizes of $\mathcal{O}(1)\km$.  This sequestration of mass dramatically reduces the
stochasticity generated by the largest bodies, which have sizes of $\mathcal{O}(1000) \km$.  

We conclude that Neptune's Brownian motion did not impede in any substantive way the planet's capture and retention of Resonant KBOs.
 
\subsection{Extensions}\label{sec-extend}
 
\subsubsection{Single Kick to Planet}

Our focus thus far has been on the regime in which many stochastic  
kicks to the planet are required for resonant particles to escape.  
Of course, a single kick from a planetesimal   
having sufficiently large mass $m_1$ could    
flush particles from resonance. To estimate $m_1$,    
we equate the change in the planet's semi-major axis from a single   
encounter, $\Delta a_{\rm p}$, to the maximum half-width  
of the resonance, $\delta a_{\rm p,lib}/2$ (see Equation [\ref{eqn-aplib}]   
and related discussion). In the likely event that the perturber's    
eccentricity $e$ is of order unity, then   
$\max (\Delta a_{\rm p}) \sim (m_1/M_{\rm p}) a_{\rm p} e$   
(Equation [\ref{eqn-delamax}]) and therefore      
$m_1 \gtrsim 0.6 \, (0.5/e) \, M_{\oplus}$ for Plutinos to escape resonance.  
Our estimate for $m_1$ agrees with that of Malhotra (1993).  
Why such enormous perturbers have not been observed today is unclear   
and casts doubt on their existence (Morbidelli, Jacob, \& Petit 2002). If such  
an Earth-mass planetesimal were present over the duration $T$ of Neptune's   
migration, then the likelihood of a resonance-destabilizing encounter  
would be $P_1 \sim \dot{\overline N}T \sim 10^{-2} (0.5/e)^4$, where  
the encounter rate $\dot{\overline N}$ is given by Equation (\ref{eqn-Ndot})  
with $\Sigma_{m} \sim m_1 / (2\pi a_{\rm p}^2)$, and we have set $T\sim 2.7\times 10^7 \yr$.                
 
\subsubsection{Kicks to Resonant Planetesimals}  

Finally, we have ignored in this work how disk planetesimals directly
perturb the semi-major axis of a resonant particle.  This neglect does
not significantly alter our conclusions.  
Take the resonant
planetesimal to resemble a typical Resonant KBO observed today, having
size $s_{\rm res} \sim 100 \km$. Then its Hill velocity is $e_{\rm
H,res}\Omega a \sim 10^{-4} \Omega a$. The relative velocity between
the resonant planetesimal and an ambient, perturbing planetesimal
greatly exceeds this Hill velocity, if only because migration in
resonant lock quickly raises the eccentricity of the resonant
planetesimal above $e_{\rm H,res}$. Equation (\ref{eqn-superrms}),
appropriate for the super-Hill regime, implies that $\langle
\dot{a}_{\rm p,rnd}^2 \rangle^{1/2} \propto M_{\rm p}^0$---the RMS
random velocity does not depend on the mass of the object being
perturbed!
Therefore when both the resonant planetesimal
and the planet are scattering planetesimals in the super-Hill regime,
their random walks are comparable in vigor.  The conservative limit of
$\mathcal{O}(100)$ km on the planetesimal size which allows resonance
retention is derived, by contrast, for the sub-Hill, maximum
stochasticity regime, and is therefore little affected by these
considerations.\footnote{The probability that a planetesimal will be ejected from resonance by a planetesimal of comparable mass in a single encounter is negligibly small.}
 
\acknowledgements
This work was made possible by grants from the National Science Foundation,
NASA, and the Alfred P.~Sloan Foundation.
We thank B.~Collins, J.~Hahn,  A.~Morbidelli, I.~Shapiro, J.~Wisdom, and        
an anonymous Protostars and Planets V referee for encouraging
and informative exchanges.
Section \ref{sec-sizedist} was written in response to insightful comments
by Mike Brown and Re'em Sari.
An anonymous referee helped to improve the presentation of this paper.

\appendix 
\section{Absorption Probability for Brownian Motion with a Double Boundary}\label{ap-abs}  
  
Here we derive Equation (\ref{eqn-abs}), the probability that a particle experiencing Brownian motion between two absorbing boundaries has not been absorbed by time $t$ (e.g., Grimmett and Stirzaker 2001b).  Consider Brownian motion along a path $x(t)$ with $x(0) = 0$ and absorbing boundaries at $x=\pm b$, $b \ge 0$.  The probability density distribution $f(x,t)$ satisfies the diffusion equation  
\begin{equation}\label{eqn-diffeq}  
\frac{\partial f}{\partial t} = \frac{1}{2} D \frac{\partial^2 f}{\partial x^2} \,\, ,  
\end{equation}  
where $D$ is the diffusion coefficient.  The absorbing boundaries generate the boundary conditions  
\begin{equation}\label{eqn-bc}  
f(\pm b, t) = 0  
\end{equation}  
for all time,\footnote{Equation (\ref{eqn-bc}) holds as long as $D$ is non-zero.  Particles near the boundary are carried across by fluctuations too quickly to maintain a non-zero density $f$ at $x=\pm b$ (see Grimmett and Stirzaker 2001a for a proof).} and the initial condition is  
\begin{equation}\label{eqn-ic}  
f(x,0) = \delta(x) \,\, .  
\end{equation}  
  
To solve for $f(x,t)$, we expand $f$ in a Fourier series, keeping only terms that satisfy (\ref{eqn-bc}):  
\begin{equation}\label{eqn-fser}  
f(x,t) = \sum_{n=1}^{\infty} k_n(t)\sin{\left(\frac{n\pi (x+b)}{2b}\right)} \,\, .  
\end{equation}  
Plugging (\ref{eqn-fser}) into (\ref{eqn-diffeq}), we find  
\begin{equation}  
k_n(t) = c_n e^{-\lambda_n t}  \,\, ,  
\end{equation}  
where the $c_n$'s are constants and   
\begin{equation}  
\lambda_n \equiv \frac{n^2\pi^2}{8b^2}D \,\, .  
\end{equation}  
The $c_n$'s must satisfy (\ref{eqn-ic}).  From Fourier analysis at time $t=0$, we find  
\begin{eqnarray}  
c_n &=& \frac{1}{2b}\int_0^{4b} \Big[\delta(y-b)-\delta(y-3b)\Big]\sin\left(\frac{n\pi y}{2b}\right) dy \\  
&=& \frac{1}{b}\sin\left(\frac{n\pi}{2}\right) \,\, .  
\end{eqnarray}  
The probability that the walker has not yet crossed either of the absorbing boundaries at time $t$ is  
\begin{eqnarray}  
P_{\rm keep}(t) &=& \displaystyle{\int_{-b}^{b} f(x,t)\,dx} \\  
 &=& \displaystyle{\int_{-b}^{b} \sum_{n=1}^{\infty} c_n e^{-\lambda_n t} \sin\left(\frac{n\pi(x+b)}{2b}\right) dx} \\  
 &=& \displaystyle{\sum_{n=1}^{\infty} \frac{4}{n\pi} \sin^3\left(\frac{n\pi}{2}\right) e^{-(n\pi)^2Dt/(8b^2)}}  \,\, .  
\end{eqnarray}


\begin{deluxetable}{cll}
\tabletypesize{\small}
\tablewidth{0pt}
\tablecaption{Frequently Used Symbols \label{tbl-symbol}}
\tablehead{
\colhead{Symbol} &
\colhead{Definition} &
\colhead{Remark}
}
\startdata
$t$ & Time & \ldots \\
$a$ & Planetesimal Semi-Major Axis & \ldots \\
$e$ & Planetesimal Eccentricity & \ldots \\
$\Omega$ & Planetesimal Orbital Angular Velocity & \ldots \\
$u$ & Planetesimal Random (Epicyclic) Velocity & $\sim$$e\Omega a$ \\
$m$ & Planetesimal Mass & \ldots \\
$s$ & Planetesimal Radius & \ldots \\
$\rho$ & Planetesimal Internal Density & $2 \gm \cm^{-3}$ \\
$a_{\rm d}$ & Mean Radius of Planetesimal Disk Annulus & \ldots \\
$\Sigma$ & Total Disk Surface Density & $0.2 \gm \cm^{-2}$ for minimum-mass\\
 & (Mass Per Unit Face-On Area)  & trans-Neptunian disk \\
$\Sigma_m$ & Disk Surface Density in Planetesimals of Mass $m$ & \ldots \\
$\mathcal{M}$ & $2\pi \Sigma_m a_{\rm d}^2 / M_{\rm p}$, Parameterizes
Surface Density & $2$ for $\Sigma_m = 0.2 \gm \cm^{-2}$, $a_{\rm d} = 26.6\AU$, \\
 & & and $M_{\rm p} = M_{\rm N}$ \\
$a_{\rm p}$ & Planet Semi-Major Axis & \ldots \\
$\Omega_{\rm p}$ & Planet Orbital Angular Velocity & \ldots \\
$M_{\rm p}$ & Planet Mass & \ldots \\
$M_\ast$ & Mass of Host Star & \ldots \\
$R_{\rm H}$ & Hill Radius of Planet $\equiv a_{\rm p}(M_{\rm p}/(3M_{\rm *}))^{1/3}$  & \ldots \\
$e_{\rm H}$ & Hill Eccentricity $\equiv R_{\rm H}/a_{\rm p}$ & \ldots \\
$v_{\rm H}$ & Hill Velocity $\equiv \Omega_{\rm p}R_{\rm H}$ & \ldots \\
$\dot{a}_{\rm p,rnd}$ & Planet Random Migration Velocity &
time-averages to zero \\
$\dot{a}_{\rm p,avg}$ & Planet Average Migration Velocity & assumed
known function \\
$x$ & $a-a_{\rm p}$ & $|x| \lesssim a_{\rm p}$ \\
$\mathcal{R}$ & Minimum Value of $|x|/R_{\rm H}$ & $\gtrsim 1$ \\
$b$ & Impact Parameter of Planet-Planetesimal Encounter & $>0$ \\
$\Delta t_{\rm e}$ & Duration of Planet-Planetesimal Encounter &
$\sim$$1/\Omega_{\rm p}$ at longest \\
$\Delta Q$ & Change in Quantity $Q$ from a Single Encounter & evaluated well before \\
& & and well after encounter, e.g., $\Delta a_{\rm p}$ \\
$\Delta t$ & Arbitrary Time Interval & \ldots \\
$N$ & Number of Planetesimals Encountered by Planet in $\Delta t$ &
Poisson deviate \\
${\overline N}$ & Mean of $N$ & \ldots \\
$\dot{\overline N}$ & Mean Rate of Planetesimal Encounters by Planet & \ldots
\\
$\left<\dot a_{\rm p,rnd}^2\right>^{1/2}$ & Root-Mean-Squared (RMS) &
$\propto 1/\sqrt{\Delta t}$ \\
 & Random Migration Velocity Over $\Delta t$ & \\
$\mathcal{C}$ & Numerical Coefficient for $\left< \dot{a}^2_{\rm
p,rnd} \right>^{1/2}$ & Equation (\ref{eqn-rosetta}), \\
& & estimated to be of order several \\
$D$ & Diffusivity of Planet's Semi-Major Axis $= (\Delta a_{\rm p})^2
\dot{\overline N}$ & \ldots \\
$a_{\rm p,i}$ & Initial Semi-Major Axis of Planet, Pre-Migration & 23.1 AU \\
$a_{\rm p,f}$ & Final Semi-Major Axis of Planet, Post-Migration & 30.1 AU \\
$\tau$ & Exponential Timescale for Migration & Equation (\ref{eqn-aavg}) \\
$T$ & Total Duration of Migration & \ldots \\
$\sigma_{a_{\rm p},T}$ & $\left< (a_{\rm p}-a_{\rm p,avg})^2
\right>^{1/2}$ After Time $T$ & $\propto T^{1/2}$, Equation
(\ref{eqn-sigmaT}) \\
$P_{\rm keep}$ & Probability a Resonant Particle is Retained &
Equation (\ref{eqn-abs}) \\
 &  in Resonance After Time $T$ & \\
$S_{\rm rnd}$ & $\int_0^t \dot{a}_{\rm p,rnd} \, dt$ & \ldots \\
$\Delta S_{\rm rnd}$ & $S_{\rm rnd}(t+\Delta t) - S_{\rm rnd}(t)$ & \ldots \\
$m_{\rm crit}$ & Maximum Planetesimal Mass Satisfying $P_{\rm keep}
\sim 1,$ & Equation (\ref{eqn-mlim}) \\
& For Disks of a Single Planetesimal Mass & \\
$s_{\rm crit}$ & Maximum Planetesimal Radius Satisfying $P_{\rm keep}
\sim 1,$ & Equation (\ref{eqn-smax}) \\
& For Disks of a Single Planetesimal Mass & \\
$f$ & Probability Density & Equation (\ref{eqn-disppdf}) \\
$M_{\rm N}$ & Mass of Neptune & \ldots \\
$e_{\rm res}$ & Eccentricity of Resonant Planetesimal & \ldots \\
$\phi$ & Resonance Angle (Libration Phase) &
Equation (\ref{eqn-resangle}) \\
&  of Resonant Planetesimal & \\
$\delta a_{\rm p,lib}$ & Maximum Width of Resonance,  & Equation (\ref{eqn-aplib})\\
& Referred to Planet's Orbit  &  and related discussion \\
$d\eta/ds$ & Differential Size Spectrum of & \ldots \\
& Noise-Generating Planetesimals & \\
$q$ & Index for Power-Law Size Distributions & $d\eta/ds \propto s^{-q}$ 
\enddata
\end{deluxetable}

\end{document}